\documentclass[12pt,preprint]{aastex}
\shorttitle{Accretion in Binary}
\shortauthors{HANAWA}

\begin{document}

\title{Gas Accretion from a Circumbinary Disk}

\author{Tomoyuki Hanawa\altaffilmark{1}, 
Yasuhiro Ochi\altaffilmark{2} \& Koichi Ando\altaffilmark{2}}
\altaffiltext{1}{Center for Frontier Science,
Chiba University, Inage-ku, Chiba, 263-8522, Japan}
\email{hanawa@cfs.chiba-u.ac.jp}
\altaffiltext{2}{Graduate School Science,
Chiba University, Inage-ku, Chiba, 263-8522, Japan}

\begin{abstract}
A new computational scheme is developed to study
gas accretion from a circumbinary disk.
The scheme decomposes the gas velocity into two 
components one of which denotes the Keplerian
rotation and the other of which does the deviation
from it.  This scheme enables us to solve the
centrifugal balance of a gas disk against gravity with
better accuracy, since the former inertia force 
cancels the gravity.
It is applied to circumbinary disk rotating around binary of which
primary and secondary has mass ratio, 1.4:0.95.
The gravity is reduced artificially softened only in small 
circular regions around the primary and secondary.  The radii
are 7 \% of the binary separation and much smaller than
those in the previous grid based simulations. 
7 Models are constructed to study dependence on the
gas temperature and the initial inner radius of the
disk.  The gas accretion shows both fast and slow 
time variations while the binary is assumed to have
a circular orbit.  The time variation is due to
oscillation of spiral arms in the circumbinary disk.
The masses of primary and secondary disks increase
while oscillating appreciably.  The mass accretion rate 
tends to be higher for the primary disk although the 
secondary disk has a higher accretion rate in certain 
periods.  The primary disk is perturbed 
intensely by the impact of gas flow so that 
the outer part is removed.  The secondary disk is 
quiet in most of time on the contrary.  Both the primary and secondary 
disks have traveling spiral waves which transfer angular
momentum within them.
\end{abstract}


\section{INTRODUCTION}

Accretion from a circumbinary disk is expected to play 
an essential role in binary formation.  Current
star formation theory assumes that stars have
only 0.01~M$_\odot$ at their birth and gain most of
the mass at the main sequence stage through gas accretion.
The accretion is mainly through disks surrounding the
stars.  When the stars are members of binaries, they
are expected to accrete gas from the circumbinary 
disks when they are young. 

Recent infrared imaging observations support the gas 
accretion from the circumbinary disk.  
The disk around CoKu Tauri/4 is of circumbinary rather than the 
transitional \citep{ireland08}.  The inner hole of the disk is 
likely to be cleared by the newly discovered binary with the
separation of $\sim$8~AU. UY~Aur has a circumbinary disk
having a clumpy structure \citep{hioki07}, which implies
dynamical interaction with the binary inside the disk.

Despite the importance, gas accretion is still mysterious 
in binary system.  The first numerical simulation of the
gas accretion onto binary was performed by \citet{artymowicz96}.
They demonstrated mass flow through gaps in circumbinary disks
by SPH simulations.  Shortly later \citet{bate97} demonstrated that
the secondary has a much higher accretion rate than the
primary when accreting
gas has a relatively large angular momentum.  Their result
seems to be plausible since the secondary has a larger
orbit and hence is closer to the inner rim of the 
circumbinary disk.  If this is the case, however, the
mass ratio should increase and approach to unity.  
Accordingly equal mass binaries should be dominant 
in contradiction to observed statistics.  Unequal mass
binaries are rather common.  According to 
\citet{duquennoy91}, the mean mass ratio is
$\sim$~0.4 for the solar type star stars.
\citet{reid97} finds that the mass ratio distribution
is approximately flat.  Although the statistic may 
suffer from observational selection bias, unequal mass
binaries are surely not rare.  Present SPH simulations
of star formation predict dominance of nearly equal mass 
binaries and do not reproduce the statistics 
\citep[see, e.g.,][]{delgado-donate05,clarke08}. 

\citet{ochi05}, which is referred as Paper I in the following,
computed gas accretion onto binary using
a grid based code and obtained qualitatively different results.
The primary has a larger accretion rate in their 
simulations.  Their simulations employ different methods 
and assumptions from those of the previous SPH simulations.  
First the SPH simulations are three dimensional,
while Paper I is two dimensional.  Second
the SPH simulations assume a cold isothermal gas 
while Paper I assumes a warm isothermal gas.
Third the method of gravity softening is different.
The SPH particles are absorbed when they enter
into  the regions close to either of the primary and secondary.
On the other hand, the gas is accumulated in the regions 
around the primary and secondary in Paper I.
The regions of artificial gravity softening are much wider
in Paper I.  The radii are 5~\% of the binary separation
in the SPH simulations while they are 20~\% of the binary
separation in Paper I.

The differences mentioned above may not account for 
the difference in the results \citep[see, e.g.,][]{clarke08}. 
The gas is concentrated on the orbital plane in the 
SPH simulations. 
Two dimensional SPH simulations assuming a warm gas
does not reproduce the gross features of Paper I, 
as shown in Figure 2 of \citet{clarke08}.  
This discrepancy should be resolved to clarify the
mystery of unequal mass binaries.

We have developed a new scheme to solve gas accretion
onto binary, since an ordinary grid-based scheme does
not work in practice.

A standard scheme demands a high spatial
resolution around the primary and secondary so that 
the grid spacing is much smaller than the scaleheight.  
The scale height around the primary is evaluated to be 
$ H _1~=~r _1 ^2  c _s ^2 / (GM) $, where $ r _1 $, 
$ c _s $, $ G $, and $ M _1 $ denote the radial distance 
from the primary, the sound
speed, the gravitational constant, and the primary mass,
respectively.  Then the required spatial resolution
is expressed as $ \Delta r \, \ll \, (c _s/v _\varphi) ^2 r $,
where $ v _\varphi $ denotes the Keplerian rotation 
velocity around the primary, $ \sqrt{GM _1/ r _1} $.
The required resolution is enormous even when the gas is
relatively warm.  Paper I studies the warm gas of 
$ c _s \, = \, 0.25 $ in the unit system in which the 
binary separation and angular velocity are unity. 
When the mass ratio is $ q~=~0.7 $,
the scaleheight is $ H _1 \, = \, 0.021 $ at $ r _1 \, = \, 0.2 $,
i.e., at the outer edge of the region of artificial gravitational
softening.  If the gas is colder or closer to the primary, the
scaleheight is shorter and higher spatial resolution is required.

We have overcome the difficulty by introducing the reference
velocity.  The reference velocity is set so that the
acceleration cancels the gravity.  The introduction of
reference velocity reduces the source term in the 
hydrodynamical equations.  As a result we have succeeded 
in reducing the softening radius down to 7~\% of the 
mean separation.

The new scheme improved solutions around the primary and
secondary.  This improvement is essential to examine the
discrepancy between the SPH simulations and Paper I.
The discrepancy is prominent within the Roche lobe.  
The accretion from the circumbinary disk is mainly through
L2 point, i.e., the side close to the secondary. The gas
streams around the secondary in Paper I while it accretes
onto the secondary in SPH simulations.  Although shock 
waves form in both types of simulations, the location
is scheme dependent.  The recipient of gas accretion depends
critically on the shock location, since the shock
energy dissipation settles the gas.  The flow is highly
supersonic in the circumstellar disks, since  
the Keplerian rotation velocity is 
$ v _\varphi~=~1.71~(r _1/0.2) ^{-1/2}  $ for $ q~=~0.7$.
The Mach number should exceed 30,
if $ c _s~=~0.05 $ as assumed in the SPH simulations.
Highly supersonic flow is not easy to handle.  A small 
numerical fluctuation may cause spurious turbulence 
and shock.  Furthermore a stationary shock produces
a strong density jump since the density enhances inversely
proportional to the square of the Mach number.
Our new scheme resolved spiral shock 
waves formed in the cicumstellar disks. 

Moreover we employed nested grids to obtain higher
spatial resolution around the primary and secondary.
The nested grids consists of rectangular grids having
different resolutions. The highest resolution of
0.12~\% binary separation is achieved in the
regions close to the primary and secondary.
On the other hand, the outer boundary is
set at 39.2 times distant from the center of gravity.
The nested grids eliminate possible wave reflection
at the outer boundary.  

We also modified initial and outer boundary 
conditions from Paper I.  The circumbinary disk is
assumed to rotate with the Keplerian velocity
and to have uniform surface density beyond the
initial inner edge at the initial stage.  
The outer boundary condition is set so that the surface
density and velocity are fixed to be constant in time.
These initial and boundary conditions reduce fluctuations
due to gas accretion.  The initial stage may correspond
to a stage at which accreting gas has been accumulated
in the circumbinary disk.

This paper is organized as follows.  Our model and numerical 
methods are summarized in \S2.  The numerical results are
shown in \S3 with emphasis on oscillation of spiral waves
in the circumbinary disk.  The oscillation results in 
time variation in the accretion rate.
We discuss implications of our results in \S4.
The technical details on the reference velocity method are
described in Appendix.

\section{Model}

We consider a binary system in which circumbinary and 
circumstellar disks are coplanar with the orbit.
The masses of the disks are neglected and the binary 
orbit is assumed to be circular for simplicity.
Furthermore we integrate the hydrodynamical equation in the
direction vertical to the orbital plane to reduce the
problem in the two dimensional.
Then the hydrodynamical equations are expressed as
\begin{equation}
\frac{\partial \Sigma}{\partial t} + \mbox{\boldmath$\nabla$} 
\left( \Sigma \mbox{\boldmath$u$} \right)  =  0  ,
\label{hydro1}
\end{equation}
\begin{equation}
\frac{\partial \mbox{\boldmath$u$}}{\partial t} + 
\left( \mbox{\boldmath$u$} \cdot \mbox{\boldmath$\nabla$} \right) 
\mbox{\boldmath$u$} +  \frac{1}{\Sigma}  
\mbox{\boldmath$\nabla$} \Pi  =  -  
\mbox{\boldmath$\nabla$} \Phi  -  \mbox{\boldmath$\Omega$} 
\times \mbox{\boldmath$u$} 
\label{hydro2}
\end{equation}
in the frame corotating with the binary.  
The symbols, $ \Sigma $ and $ \Pi $,
denote the surface density and vertically integrated pressure,
respectively.
The symbols, $ \mbox{\boldmath$u$}$ and $ \mbox{\boldmath$\Omega$}$,
denote the velocity in the corotating frame and angular velocity
of the binary, respectively.  The center of gravity is located
at the origin, $ \mbox{\boldmath$r$}=0$.
The gravitational potential ($ \Phi $)
at the position
$ \mbox{\boldmath$r$}$ is expressed as
\begin{equation}
\Phi = -  \sum _{k=1} ^2 \frac{G M _k}
{|\mbox{\boldmath$r$} -  \mbox{\boldmath$r$} _k|} - 
\frac{1}{2} \left( \mbox{\boldmath$\Omega$} \times 
\mbox{\boldmath$r$} \right) ^2 ,
\end{equation}
where $ M _k $ and $ \mbox{\boldmath$r _k$} $ denote
the mass and position of the $ k $-th component star,
respectively.  For simplicity we assume that the gas is 
isothermal and
\begin{equation}
\Pi = c _s ^2 \Sigma ,
\end{equation}
where $ c _s $ denotes the sound speed.  It is set to be
either $ c _s = 0.22 $ or 0.25 in this paper.

In the following we use the non-dimensional unit system in which
the unit length is the binary separation, 
$ | \mbox{\boldmath$r$} _1 - \mbox{\boldmath$r$} _2 |$,
and the unit time is the inverse of the angular velocity,
$ 1/| \mbox{\boldmath$\Omega$} | $. The gravitational constant,
$ G $, is taken to be unit for simplicity.  Then the total mass
of binary is also unity, $ M _1 + M _2 = 1 $, in this unit system.
The orbital plane is assumed to coincide with $ z = 0 $ in the Cartesian
coordinates. 
The primary is
assumed to located at $\mbox{\boldmath$r$}_1$=$(M_2,0)$, while
the secondary at $\mbox{\boldmath$r$}_2$=$(-M_1,0)$.

We decompose the gas velocity into two components,
given ($ \mbox{\boldmath$w$}$) and 
unknown ($ \mbox{\boldmath$v$}$), 
\begin{equation}
\mbox{\boldmath$u$} (\mbox{\boldmath$r$},t) =
\mbox{\boldmath$w$} (\mbox{\boldmath$r$}) +
\mbox{\boldmath$v$} (\mbox{\boldmath$r$},t) .
\label{reference}
\end{equation}
The former ($ \mbox{\boldmath$w$}$) is called reference
velocity in the following.  Substituting Equation (\ref{reference})
into Equation (\ref{hydro2}), we obtain
\begin{equation}
\frac{\partial \mbox{\boldmath$v$}}{\partial t} + 
\left( \mbox{\boldmath$u$} \cdot \mbox{\boldmath$\nabla$} \right) 
\mbox{\boldmath$v$} +  \frac{1}{\Sigma}  
\mbox{\boldmath$\nabla$} \Pi  =  \mbox{\boldmath$g$} _{\rm eff} 
-  2 \mbox{\boldmath$\Omega$} \times \mbox{\boldmath$v$} -
(\mbox{\boldmath$v$} \cdot \mbox{\boldmath$\nabla$}) 
\mbox{\boldmath$w$} , 
\label{hydro2b}
\end{equation}
where
\begin{equation}
\mbox{\boldmath$g$} _{\rm eff} = - \mbox{\boldmath$\nabla$}
\Phi - \left( \mbox{\boldmath$w$} \cdot \mbox{\boldmath$\nabla$} \right) 
\mbox{\boldmath$w$} 
- 2 \mbox{\boldmath$\Omega$} \times \mbox{\boldmath$w$} .
\end{equation}
The reference velocity is chosen so that the effective gravity,
$ \mbox{\boldmath$g$} _{\rm eff} $, is small everywhere in the
computation domain.  Thus it is nearly the Keplerian rotation 
around the each star in the vicinity while it is the Keplerian
rotation around the center of the gravity in the circumbinary disk.
The gravitational potential, $ \Phi $,
is softened only in small areas around the primary and secondary,
i.e., in the regions of 
$ | \mbox{\boldmath$r$} - \mbox{\boldmath$r$} _i | < \beta _i $.
The softening radius is taken to be 
$ \beta _1$~=~$\beta _2 $~=~0.07 in most models.

Note that the right hand side of Equation (\ref{hydro2b}) does not 
contain any derivative of $ \mbox{\boldmath$v$}$.  
Thus we integrate Equation (\ref{hydro2b})
coupled with Equation (\ref{hydro1}) numerically by an ordinary method.
Although the residual velocity ($ \mbox{\boldmath$v$}$) is
small, the nonlinear term is taken into account.  The introduction
of the reference velocity is not approximation but purely
mathematical transformation.
Further details on the numerical integration are given in the
Appendix.

The hydrodynamical equations are integrated numerically on
the nested grid consisting of ten rectangular grids.
The rectangular grids contain 512$^2$ square cells each.
Two of them have the grid spacing of 
$ \Delta x $~=~$ 1.2 \times 10^{-3} $ and cover the
vicinities of the primary and secondary.  Another
rectangular grid has the grid spacing of 
$ \Delta x $~=~0.1536 and covers the whole computation
region of $ 78.6^2 $ square area.  
The rest grids have intermediate spatial
resolutions and cover a part of the computation region.
They are arranged so that the spatial resolution is higher
in the regions closer to the primary and secondary. 

The outer boundary is placed at $ r $~=~38.9.
The surface density and velocity are replaced with the initial
values at each time step outside the outer boundary.
The outer boundary is so far from the binary that it has
no appreciable effects on the simulations. 

The initial surface density is taken to be
\begin{equation}
\Sigma \; = \; \frac{\Sigma _{\rm out} \, + \, \Sigma _{\rm in}}{2}
\; + \; 
\frac{\Sigma _{\rm out} \, - \, \Sigma _{\rm in}}{2} \,
\tanh \, \left( \frac{r \, - \, r _0}{h} \right) \; ,
\end{equation}
where $ \Sigma _{\rm out} \, = \, 1.0 $, 
$ \Sigma _{\rm in} \, = \, 0.1 $,
and $ h \; = \; 0.144 $.  Thus the initial circumbinary
disk has an inner edge at $ r \, = \, r _0 $.
The initial velocity is set so that the residual velocity,
$ \mbox{\boldmath$v$} $, vanishes.

\section{RESULTS}

This paper shows 6 models having different $ c _s $,  
$ r _0$, and $ \beta _1~(=~\beta _2)$.  The model
parameters are listed in Table~\ref{model}.
First we introduce model 1 because it has the smallest
$ c _s $ and a modestly large $ r _0 $.  
When $ c _s $ is smaller (i.e., the gas is colder), 
the disk is geometrically thinner and our 2D approximation 
is better.  When $ r _0 $ is larger, the cirucmbinary disk 
has a larger specific angular momentum and the gas can 
accrete onto the binary only after substantial angular
momentum transfer.  

\subsection{Model 1}

\begin{figure}[!hp]
\plottwo{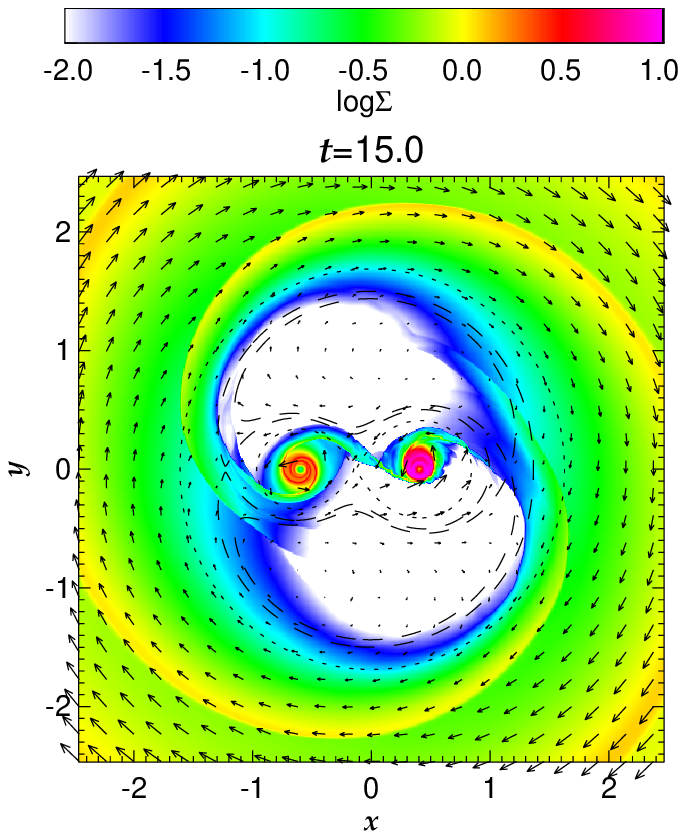}{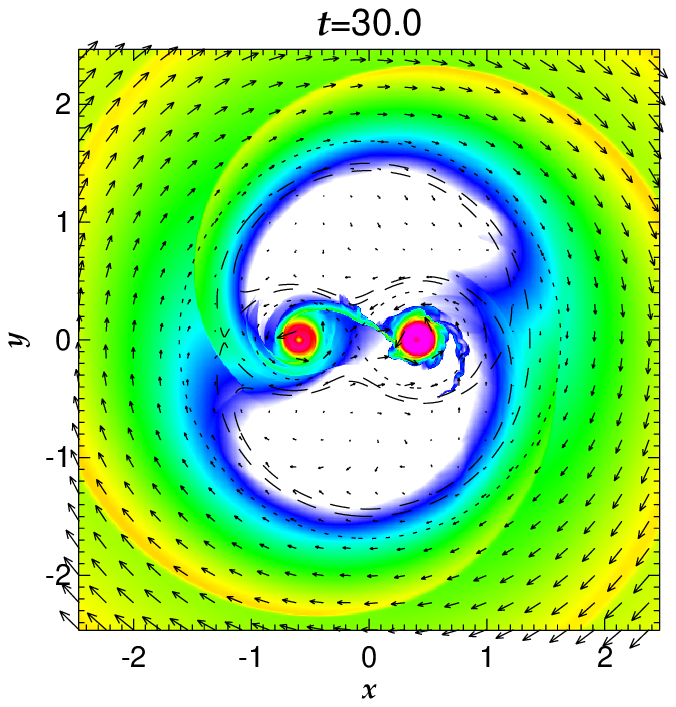}
\plottwo{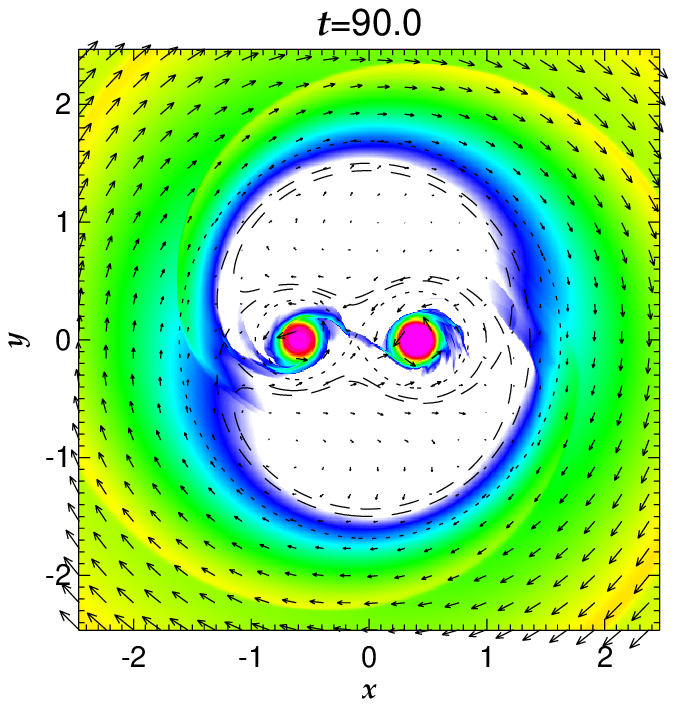}{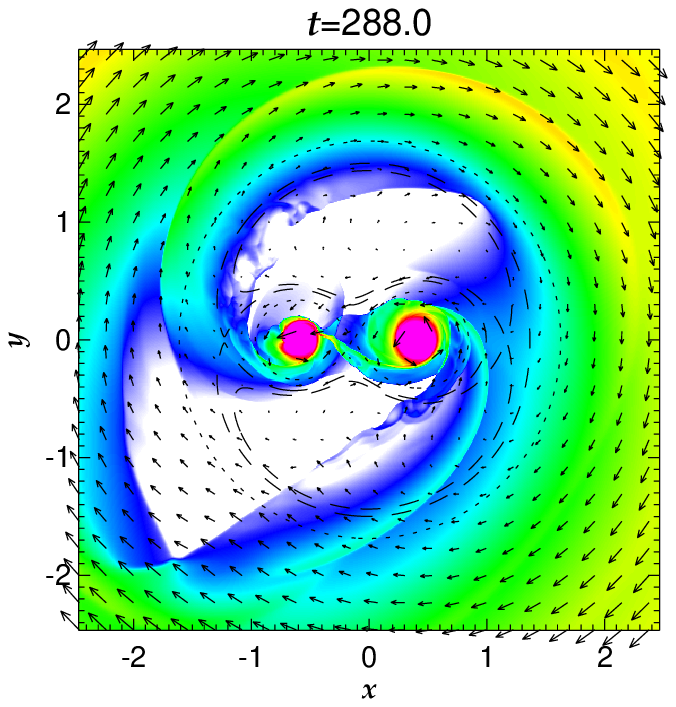}
\caption{Each panel shows the surface density (color) and
velocity distributions at $ t $~=~15, 30, 120, and 240 
in model 1.  The dashed curves are the contours of
the Roche potential.\label{model1}}
\end{figure}

The circumbinary disk has the inner edge at $ r \, = \, 2.5 $
at the initial stage of model 1.  
Figure~\ref{model1} shows the surface density distribution
at $ t $~=~15, 30, 90, and 288 in model 1.  The color
denotes the surface density in logarithmic scale while
the arrows denote the velocity in the frame corotating
with binary, $ \mbox{\boldmath$u$}$.  The primary (right)
has a more massive circumstellar disk than the secondary
(left).  

The circumstellar disks accretes gas mainly
through L2 point from the circumbinary disk.  
This is reasonable since the Roche potential is lower
at L2 point by $ \Delta \Phi \, = \, 6.70 \times 10^{-2}$.
The potential difference is appreciably larger than the
square of the sound speed, $ c _s ^2 \, = \, 4.84 \times 10 ^{-2} $.
The circumbinary disk has two-armed spiral shock waves 
(denoted by yellow in Fig.~\ref{model1}).   

The gas velocity is small near L2 point.  We surmise that
it is regulated to be the sound speed, since the Roche
potential has a saddle point at L2.  Then L2 point 
should be similar to the throat of Laval nozzle.
As well known, the gas velocity coincides with the
sound speed at the throat.  Similarly the gas velocity
should be regulated at the sound speed at L2 point.

\begin{figure}[!hp]
\centering
\plottwo{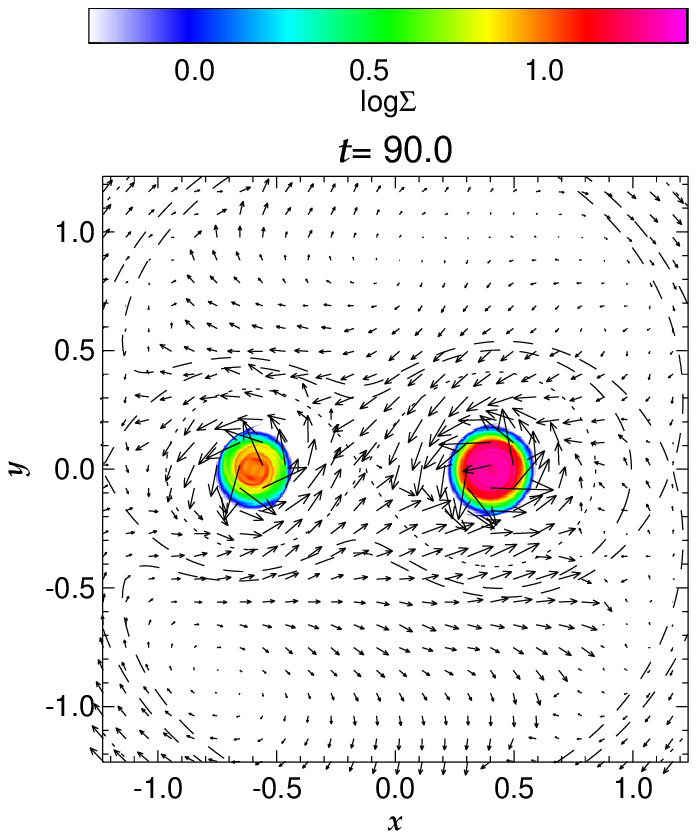}{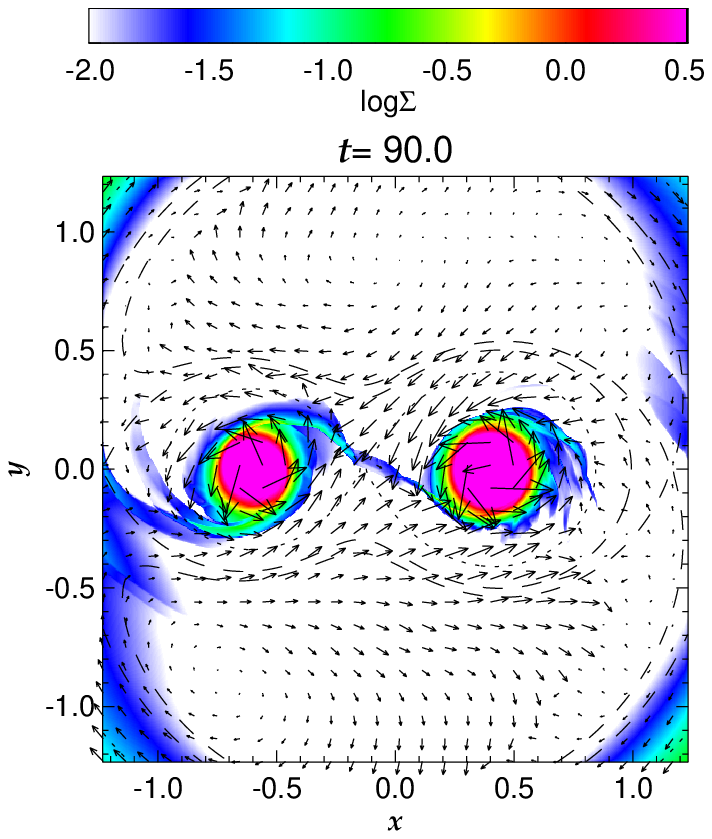}
\caption{The surface density (color) and velocity (arrows) 
distributions inside the Roche lobe at $ t $~=~90.0 in model 1.  
The dashed curves are the contours of the Roche potential.
The left and right panels denote the same stage but in the
different color scales.
\label{model1c}}
\end{figure}

\begin{figure}[!hp]
\centering
\plottwo{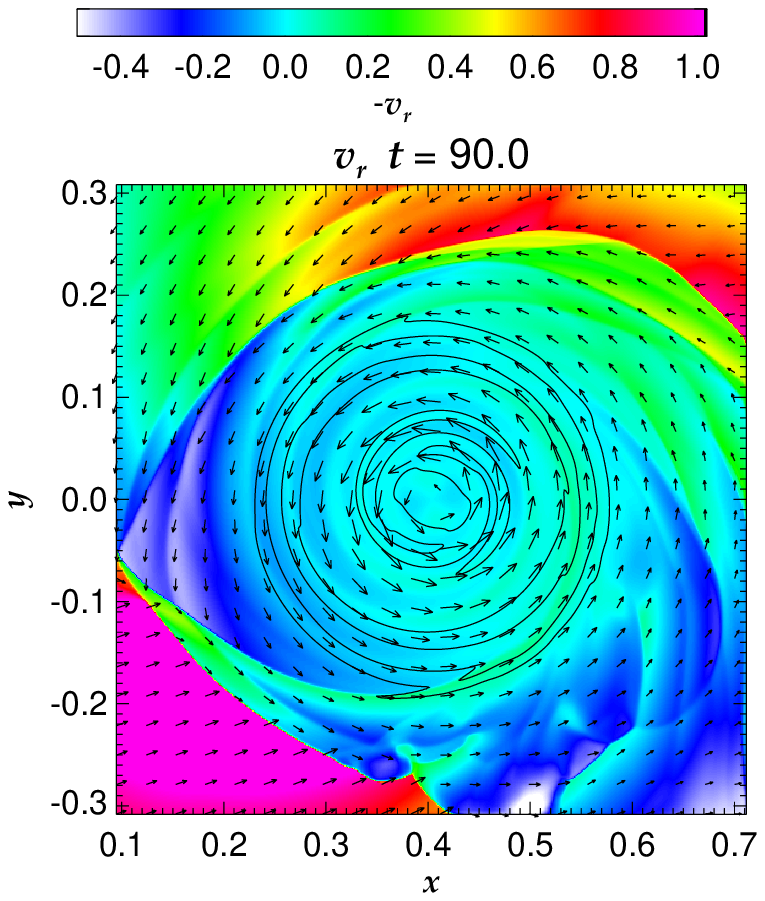}{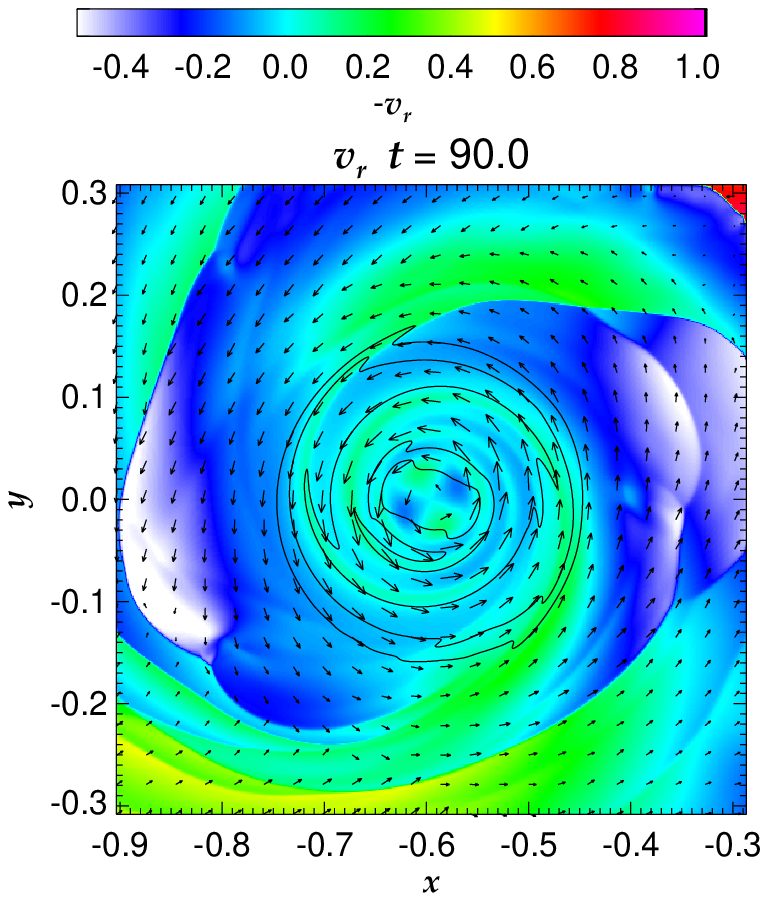}
\caption{The color denotes the radial velocity 
toward each star at $ t $~=~90.0 in model 1.
The left panel denotes the velocity toward the primary,
while the right does that toward the secondary.
\label{model1c-v}}
\end{figure}

\begin{figure}[!hp]
\begin{center}
\includegraphics[width=0.6\textwidth]{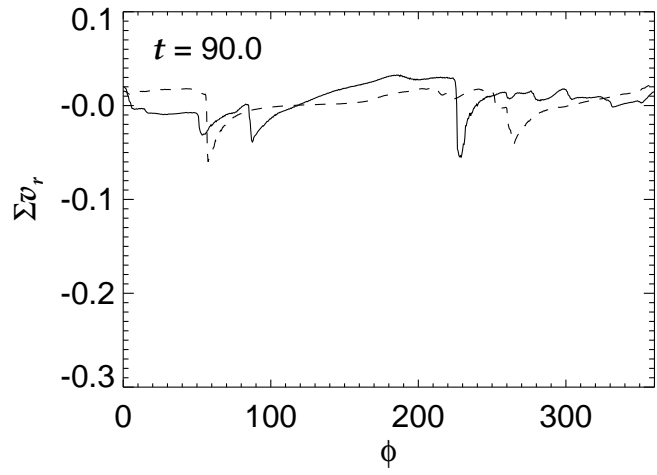}
\end{center}
\caption{The solid curve denotes the radial mass flux, 
$ \Sigma v _{r1} $, onto the primary disk as a function
of the azimuthal angle, 
$ \phi _{1} \, = \, \tan ^{-1} [y/(x \, - \, x_{1})] $,
where $ v _{r1} $ denotes the radial velocity with respect
to the primary.  The dashed curve is the same as the
solid one but for the secondary. 
\label{model1-25-phi}}
\end{figure}

Figure~\ref{model1c} is a close up view of model 1 at
$ t $~=~90.0.  The left panel is focused on the regions
of high surface density while the right is on the
regions of low surface density.  Spiral shock waves
are excited also in the circumstellar disks.
The outer part of the primary disk is
disturbed appreciably while the secondary is not.
The disturbance is likely due to the impact of gas 
flow from L2.  The gas stream is nearly normal to
to the primary disk while it is 
nearly tangential to the secondary disk.
See Figure~\ref{model1c-v} for the radial velocity
toward each star.  The primary disk receives a larger 
impact, which produces \lq a hot spot' on the 
side closer to L1 point.  The radial gas inflow is
shown as a function of the azimuthal angle centered
on the primary in Figure~\ref{model1-25-phi}.  The ordinate
denotes the mass flux, i.e., 
the product of the surface density and the
radial velocity toward the primary at $ t $~=~90.0 and
$ | \mbox{\boldmath$r$} \, - \, \mbox{\boldmath$r$} _1 | 
\, = \, 0.23$.
The azimuthal angle is $ \phi $~=~0 and $ \pi/2$ in the 
directions of increasing $ x $ and $ y $, respectively.

\begin{figure}[!hp]
\centering
\includegraphics[width=0.6\textwidth]{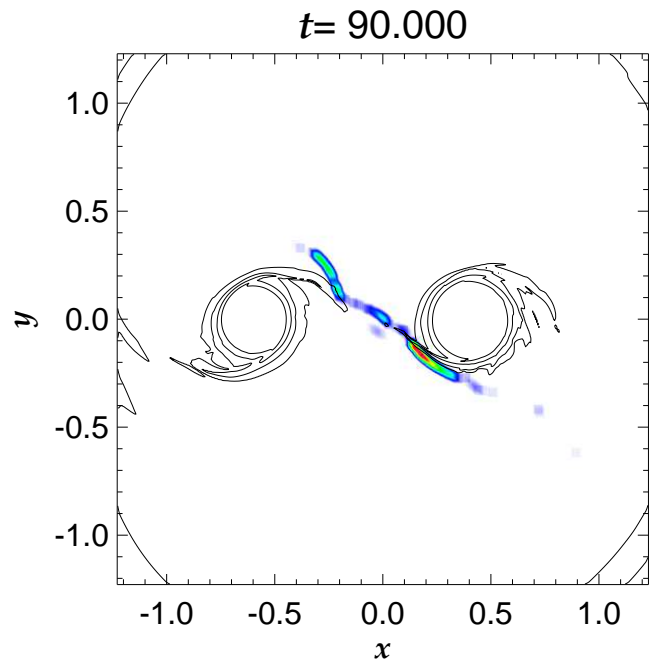}
\caption{The region of strong shock compression is
marked by color on the surface density map at
$ t $~=~90 in model 1.  The colored region is
identified by the amplitude of 
$ \mbox{\boldmath$\nabla$} \cdot \mbox{\boldmath$u$}$.
It is considerably broadened artificially, since the
the shock wave is very sharply resolved and is hard
to be recognized in the raw data.
\label{model1-st300-sh}}
\end{figure}

Figure~\ref{model1-st300-sh} shows the region of
large $ \mbox{\boldmath$\nabla$}\cdot\mbox{\boldmath$u$}$
at $ t $~=~90.0 by color.  The colored is the shock front,
since the gas is compressed strongly there.
It is clearly shown that the primary disk is associated
with the strong shock while the secondary disk is not.
Furthermore the gas stream changes its
density and velocity semi-periodically on a short timescale.  
The impact on the primary disk varies accordingly. 

Note that Figure~\ref{model1c} is different from both
the cold and warm SPH models of Delgado-Donate shown
in Figure~2 of \citet{clarke08}.  The secondary is
surrounded by a shock wave in his cold SPH model 
while no shock wave is seen in his warm SPH model. 
The shock location is closely related to the mass
accretion rate of each component star.  It is crucial
whether the gas streaming from L1 point causes a shock 
around the secondary. 

\begin{figure}[!hp]
\begin{center}
\includegraphics[width=0.6\textwidth]{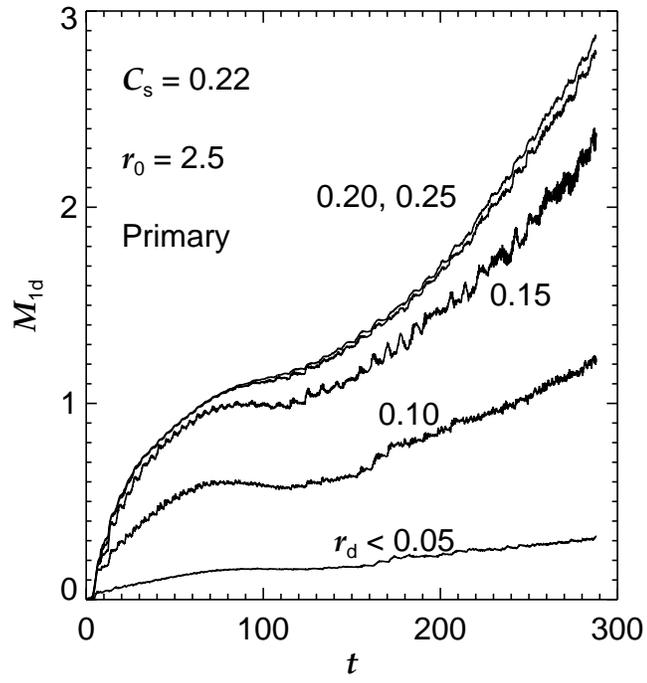}
\end{center}
\caption{Each curve denotes the total mass of gas inside 
a circle around the primary as a function of time.
The numbers denote the radii of the circles.
\label{model1-M1}}
\end{figure}

The primary disk mass increases while oscillating appreciably.
Figure~\ref{model1-M1} shows the primary disk mass 
as a function of time.  Each curve denotes the total 
mass of gas enclosed in a circle around the primary.
It also indicates that the surface density falls off
sharply around 
$ |\mbox{\boldmath$r$} \, -  \, \mbox{\boldmath$r$} _1 | 
\, = \, 0.2 $.  Thus we regard the total mass enclosed
in the circle of $ |\mbox{\boldmath$r$} \, -  \, \mbox{\boldmath$r$} _1 | 
\, = \, 0.2 $ as the disk mass in the following.

\begin{figure}[!hp]
\centering
\includegraphics[width=0.6\textwidth]{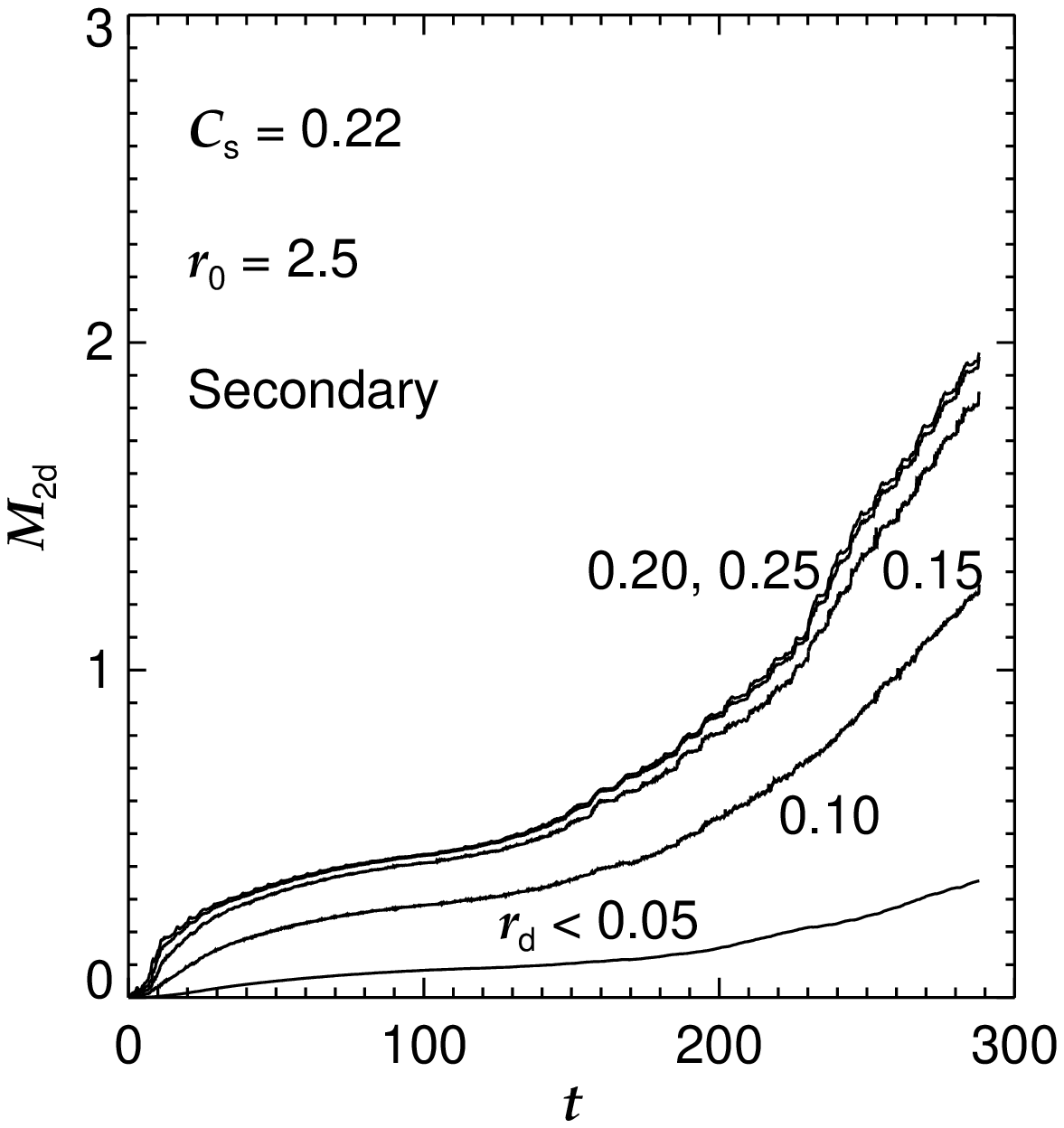}
\caption{The same as Fig.~\ref{model1-M1} but for the
secondary. \label{model1-M2}}
\end{figure}

Figure~\ref{model1-M2} is the same as Figure~\ref{model1-M1}
but for the secondary disk.  The secondary disk mass is 
smaller than the primary one and increases steadily.  
The gas flow around the secondary is smooth and the
secondary disk accretes gas without significant 
disturbance.   

\begin{figure}[!hp]
\centering
\includegraphics[width=0.6\textwidth]{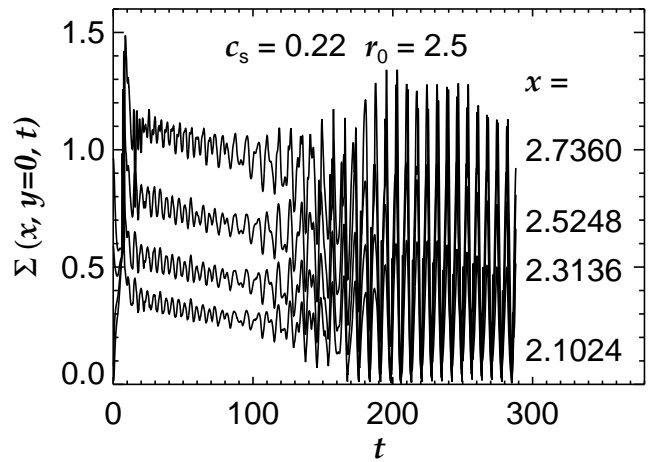}
\caption{The time variation of the surface density
on the axis of $ y $~=~0 in model 1.
\label{model1-wavex}}
\end{figure}

Semi-periodic change is also seen in the circumbinary disk.
Figure~\ref{model1-wavex} shows the time variation in
the surface density measured at four points 
on the axis of $ y \, = \, 0 $.
The period of surface density variation is about
$ P~\simeq~4.4$ when measured at a given point in the
corotation frame.  Variation is also seen in the radial
velocity.  The variation has the temporal and azimuthal 
dependence of $ \cos \, [2 (\phi \, - \, \omega t)]$ 
in the corotation frame, where the angular frequency is 
approximately $ \omega \, \simeq \, 0.7 \Omega$ in the
period $ 20 \, < \, t \, < 80 $.
Around $ t \, \simeq \, 100 $, the surface density variation
is not sinusoidal; two waves having similar but different
angular frequencies seem to be excited.
The amplitude of the wave varies on a longer timescale 
as shown in Figure~\ref{model1-wavex}.
It is large in the period $ t \, \ga \, 130 $.
This large variation is due to one armed low surface
density region shown in the lower right panel
of Figure~\ref{model1}. 

This quasi-periodic change has not been reported in the
literature.  It may be missed or smeared out in the 
previous simulations by some reasons.  
Paper I and some other simulations assumed continuous 
dynamical accretion onto the circumbinary disk, which
grows the disk and may hinder our perception.
Although \citet{guenther02} and \citet{guenther04} 
assumed quasi-static initial and boundary conditions,
they employed a large empirical viscosity and spatial
resolution much lower than ours. The numerical viscosity
and relatively low spatial resolution may damp down
the quasi-periodic change. We have learned that
recent SPH simulations have shown similar variation from
\cite{clarke09}. 

Figure~\ref{model1c} shows a thin stripe of high surface density 
connecting primary and secondary disks as in Paper I.
The stripe is formed by the collision of the gas flows rotating around 
the primary and secondary.  In practice, it is the boundary
between the domains in one of which the gas flows from
secondary to primary and in the other of which it flows
vice versa.  It is located around L1 point but
its location varies with time.  When the inflow from L2 
has a higher surface density, the stripe is closer to 
the primary.  The location of the stripe is critical
to which component of the binary accretes more.

The sum of the disk masses amounts to
$ M _{\rm 1d} \, + \, M _{\rm 2d}$~=~1.78 at $ t $~=~144.
This is equal to the mass of the circumbinary ring
of $ \Delta r $~=~0.11 at $ r \, = \, r _0 $.
The total disk mass amounts to 
$ M _{\rm 1d} \, + \, M _{\rm 2d}$~=~4.75 at $ t $~=~288.

\begin{figure}[!hp]
\centering
\includegraphics[width=0.6\textwidth]{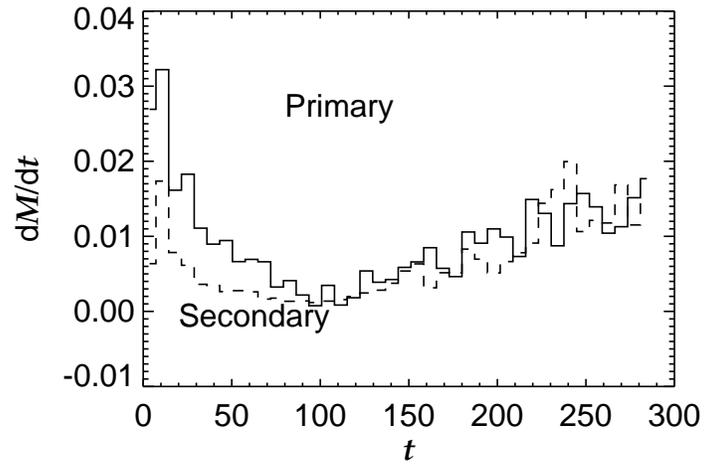}
\caption{The accretion rates of the primary and
secondary derived from the best fit polynomials to
their disk masses. \label{model1-dMdt}}
\end{figure}

The time derivative of a disk mass gives us the accretion
rate in principle.  It is, however, highly variable and hard
to read gradual change in the accretion rate.  To erase
out the short time variability, we obtain the average accretion 
rate for time intervals of $ \Delta t $~=~2.88.
Figure~\ref{model1-dMdt} shows the average accretion
rates for the primary (solid), $ {\rm d}M_{\rm 1d}/{\rm d}t$ and 
secondary (dashed), $ {\rm d}M_{\rm 2d}/{\rm d}t $. 
The accretion rate is very high in the first a few rotation
period ($ t \, \la \, 25 )$. 
The primary has a higher accretion rate than the
secondary except around  $ t \, \simeq \, 230 $ and
270.

It should be noted that there is correlation between
the accretion rates and oscillation in the circumbinary
disk.  The accretion rates are low when the circumbinary 
disk is relatively quiet, i.e., around $ t~\simeq~100 $.
This indicates that the oscillation
enhances the accretion from the circumbinary disk 
through angular momentum exchange.

\begin{figure}[!hp]
\centering
\includegraphics[width=0.6\textwidth]{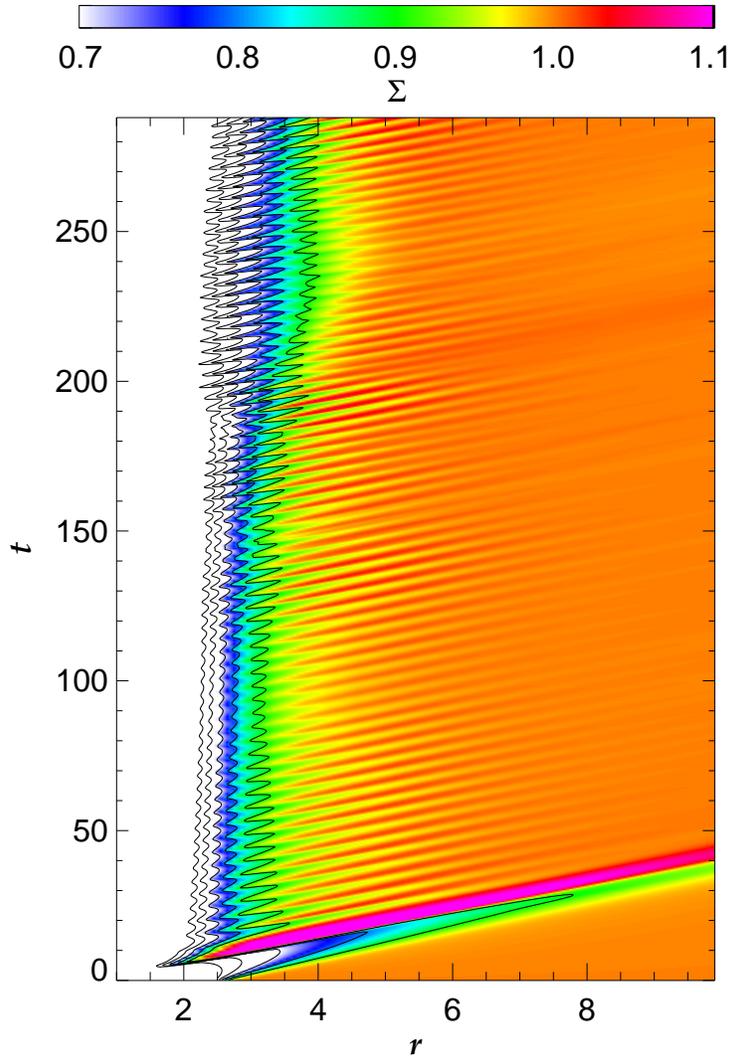}
\caption{The evolution of the azimuthally averaged
surface density of the circumbinary disk in model 1.
\label{model1-cb}}
\end{figure}

Figure~\ref{model1-cb} shows the azimuthally averaged 
surface density in the circumbinary disk as a function
of $r$ and $t$.  It shows repeatedly excited waves 
propagating outwards of which phase velocity is 
$ dr/dt \, \simeq \, 0.25 $ and 0.24 at $ r $~=~3.2 
and 4.2, respectively.  The inner edge of the circumbinary
disk retreats very slowly.  Figure~\ref{model1-cb} shows
that the amplitude of the wave varies on the timescale
of several tens rotation period.

\begin{figure}[!hp]
\centering
\includegraphics[width=0.6\textwidth]{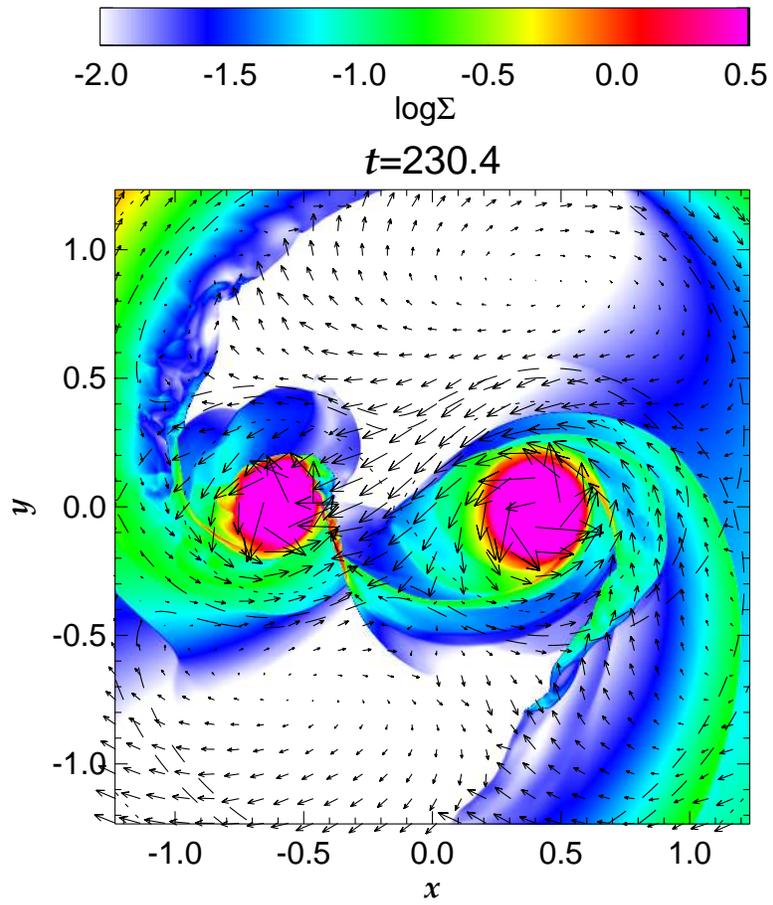}
\caption{The same as the left panel of Fig.~\ref{model1c}
but for the stage of $ t $~=~230.4.
\label{model1-st640}}
\end{figure}

At some epochs, the secondary has a higher accretion rate
than the primary. Figure \ref{model1-st640} shows the
flow at $ t $~=~230.4 as an example of such a stage.
As shown in the figure, the outer part of the circuprimary
disk is shed to flow into the secondary lobe.  The
circumsecondary disk receives the flow and has a
hot spot on the L1 point side.  Similar mass transfer
is also seen around $ t \, \simeq \, 150 $ and 190.
The inflow is still mainly through L2 point and
the inflow through L3 point is  minor.
Note that the gas flows from primary to secondary around
L1 point in this period.  Shortly after this stage,
the circumsecondary disk is disturbed very much and
captures a gas flow from L2 point.
The gas flow inside the
Roche lobe is crucial for the branching of the inflow 
from the circumbinary disk.

\subsection{Model 2}

We compare model 2 with model 1 to examine the dependence
on $ r _0 $, the initial inner radius of the cicumbinary
disk. The model parameters of models 2 are the same
as those of model 1 except $ r _0 $.

\begin{figure}[!hp]
\centering
\includegraphics[width=0.6\textwidth]{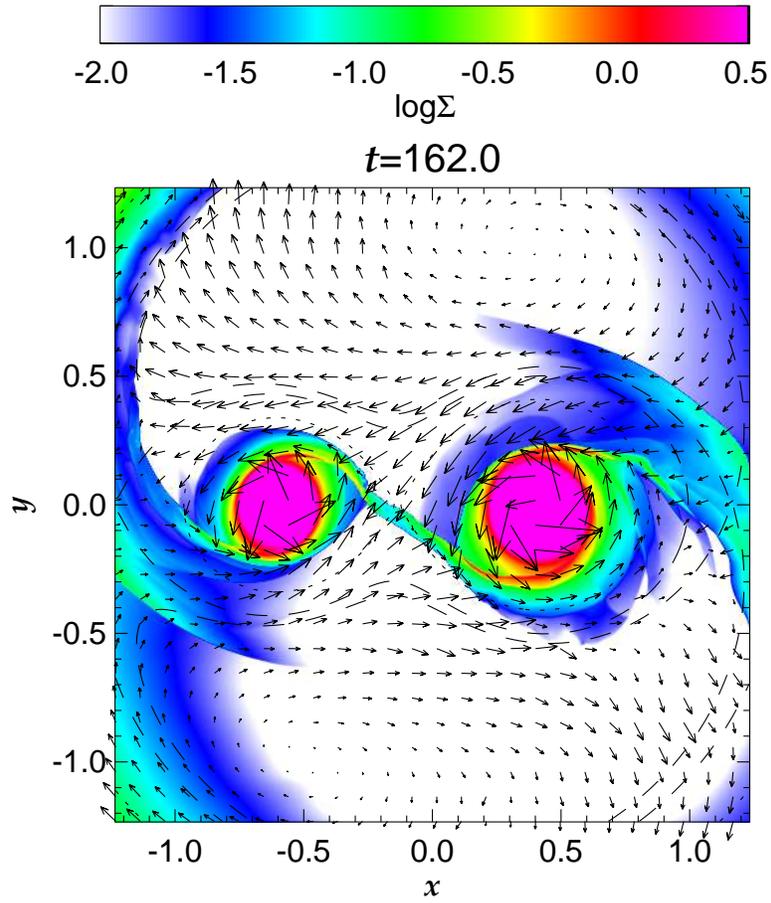}
\caption{The surface density distribution around
the binary at $ t $~=~162.0 in model 2. \label{model2}}
\end{figure}

\begin{figure}
\centering
\includegraphics[width=0.6\textwidth]{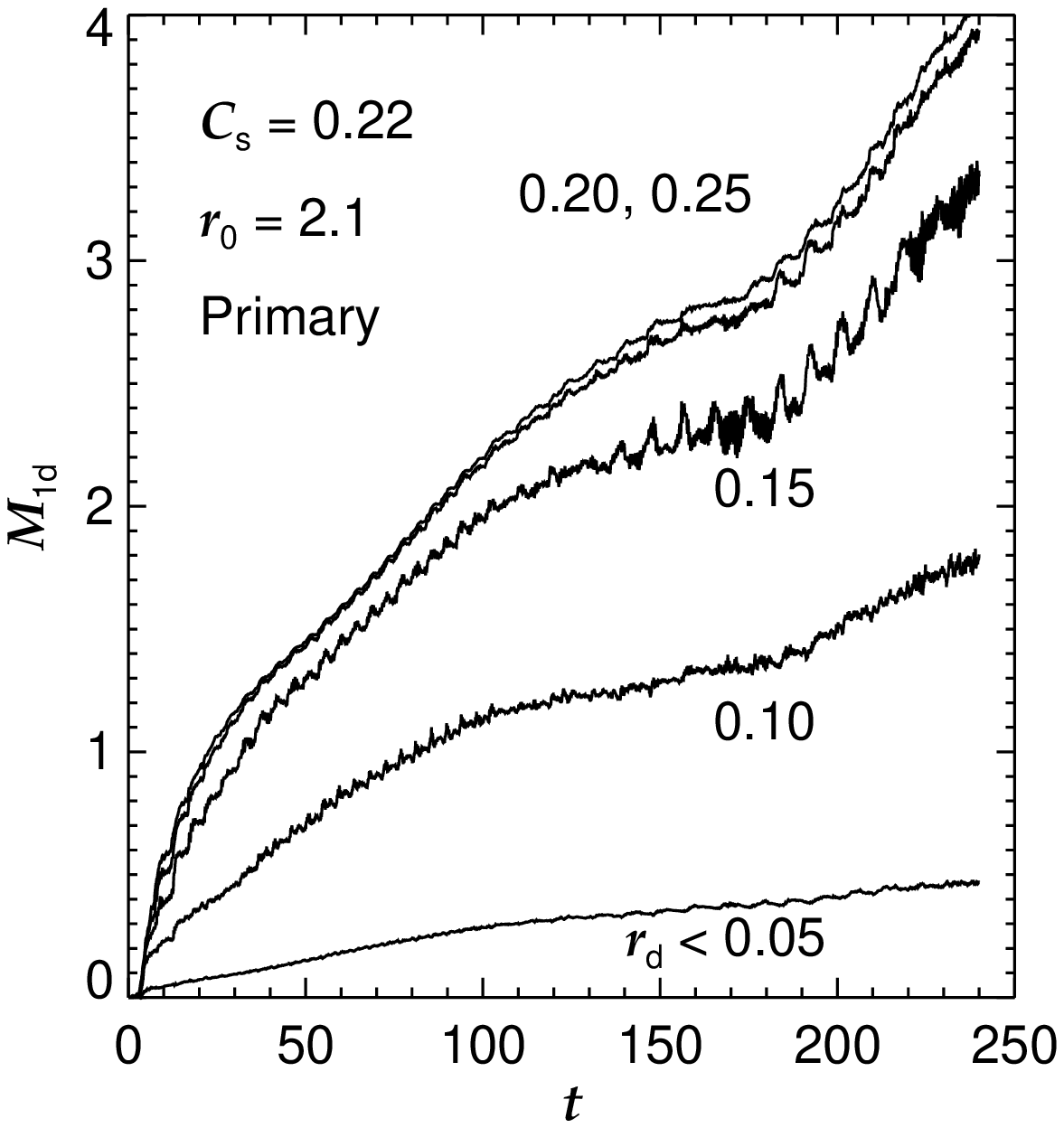}
\caption{The same as Fig.~\ref{model1-M1} but for
model 2. \label{model2-M1}}
\end{figure}

\begin{figure}
\centering
\includegraphics[width=0.6\textwidth]{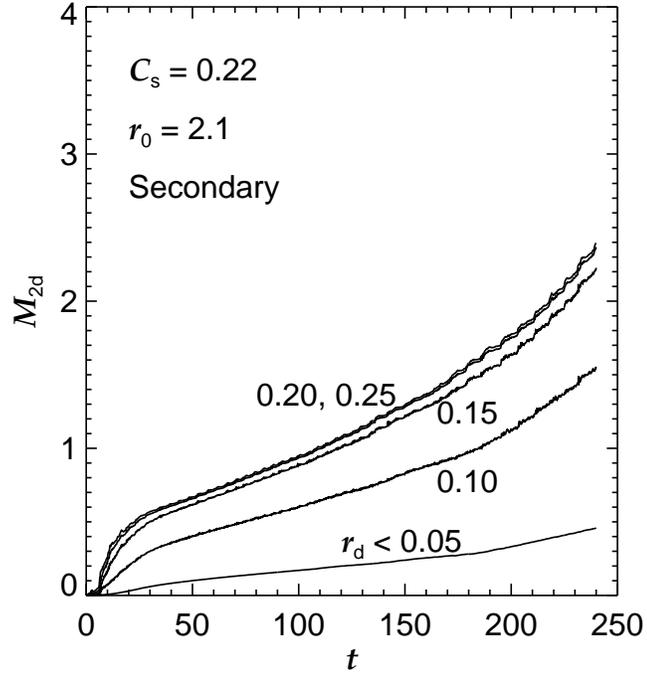}
\caption{The same ad Fig.~\ref{model1-M2} but for 
the secondary disk in model 2.
\label{model2-M2}}
\end{figure}

\begin{figure}
\centering
\includegraphics[width=0.6\textwidth]{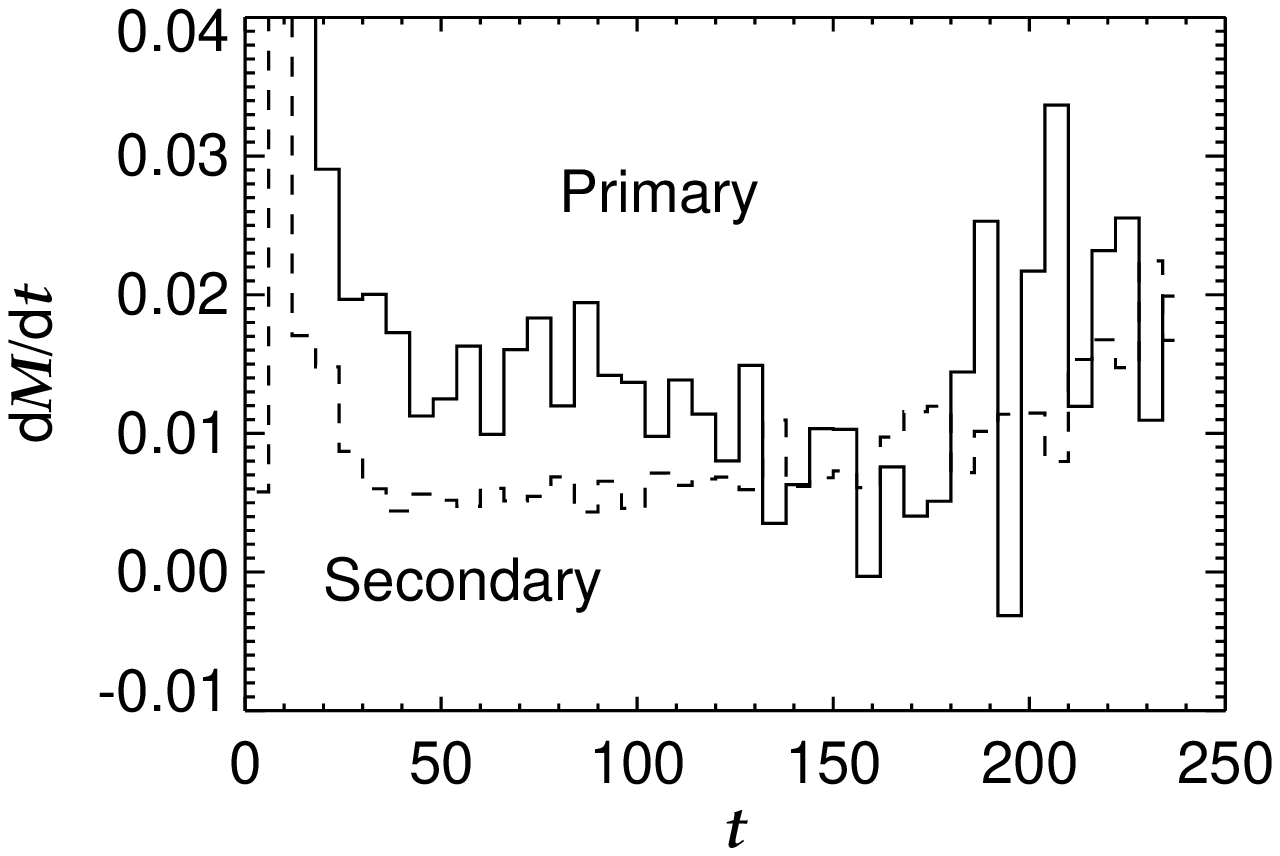}
\caption{The same ad Fig.~\ref{model1-dMdt} but model 2.
\label{model2-dMdt}}
\end{figure}

Figure~\ref{model2} shows the surface density distributions at $ t $~=~162. 
The circumstellar disk is more massive in model 2 since 
the circumbinary disk has a smaller inner radius.
The primary disk is more massive than the secondary one 
also in model 2.  The primary disk mass increases while
oscillating appreciably as shown in Figure~\ref{model2-M1}.
The secondary disk mass increases slowly and smoothly
as shown in Figure~\ref{model2-M2}.
The averaged accretion rate is shown in 
Figure~\ref{model2-dMdt}.

\subsection{Models 3, 4, and 5}

We compare model 4 with model 1 to examine the dependence
on the gas temperature.  The gas temperature is 29.1\% 
higher in model 3 than in model 1 since the isothermal
sound speed is set to be $ c _s$~=~0.25.
It is still lower than the the potential gap between L2 and L3 
points.  Thus the flow is relatively cold although it
is quite higher than that assumed in SPH simulations.

\begin{figure}
\centering
\plottwo{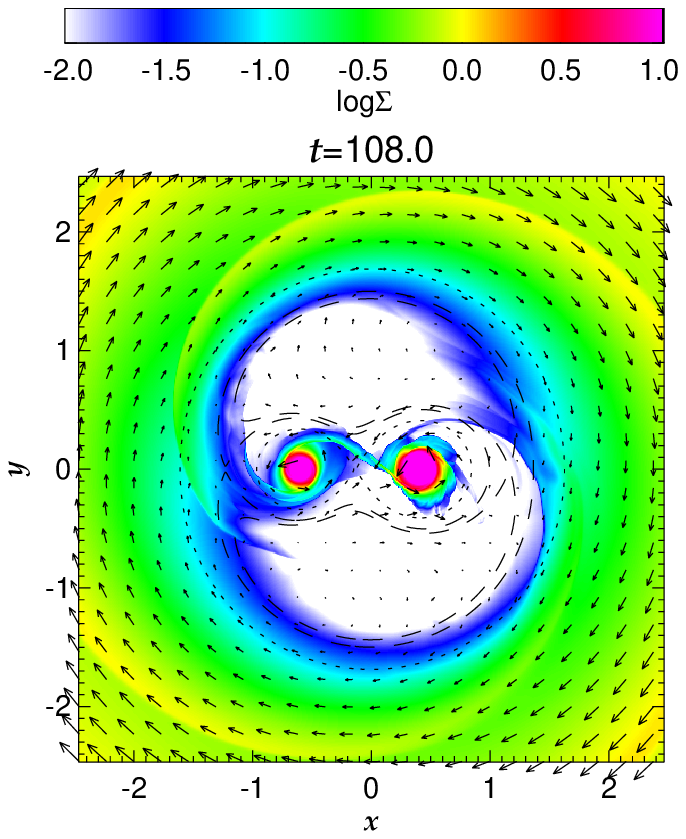}{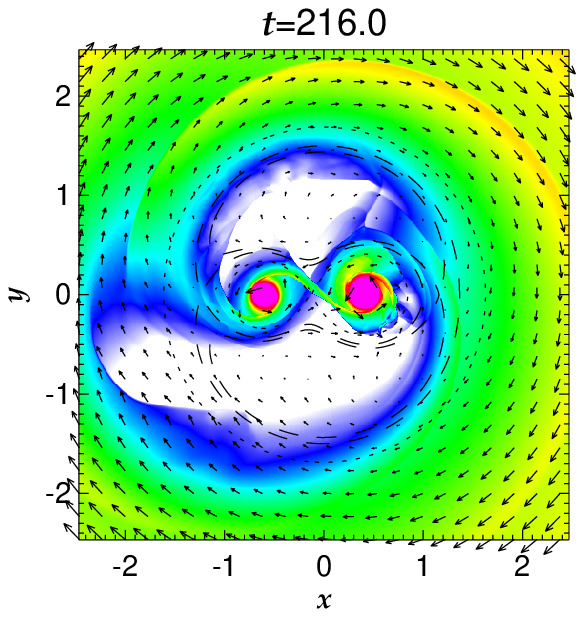}
\caption{The surface density (color) and velocity (arrows)
distributions at $ t $~=~108 and 216 in model 3.
\label{model3}}
\end{figure}

Figure~\ref{model3} shows the distributions of 
$ \Sigma $ and $ \mbox{\boldmath$u$} $ at $ t $~=~108
and 216 in model 3.  The spiral arms have a little wider 
opening angle in the circumbinary disk of model 3 than
in that of model 1. The gas inflows mainly from L2 point
and the surface density is low around L3 point. 
Remember that the potential difference is 
$ \Delta \Phi~=~6.70\times10^{-2} $ between L2 and L3.
It is larger than $ c _s ^2 $~=~$ 6.25 \times 10 ^{-2} $
also in models 3, 4, and 5.

\begin{figure}
\centering
\includegraphics[width=0.6\textwidth]{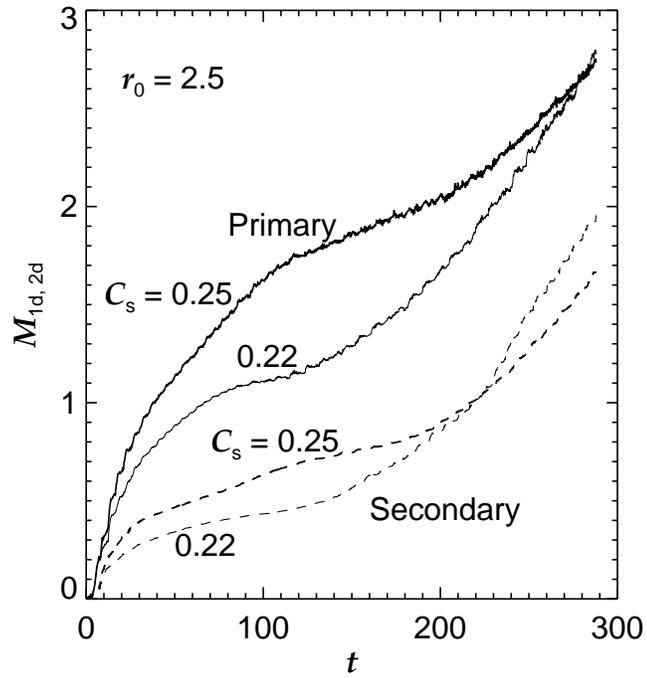}
\caption{The comparison of disk masses between model 1 (thin) 
and 3 (thick).  The solid curves denote the primary disk mass
while the dashed do the secondary one. \label{Cs-mass}}
\end{figure}

Figure~\ref{Cs-mass} compares the disk masses of model 3 with
those of model 1. Both the primary and secondary disk masses
are $\sim$50 \% larger in model 4.  This is an envidence
that the oscillating spiral arms are responsible for
the accretion from the circumbinary disk.  When $ c _s $
is higher, the angular momentum exchange should be more
effective. 

\begin{figure}
\centering
\includegraphics[width=0.6\textwidth]{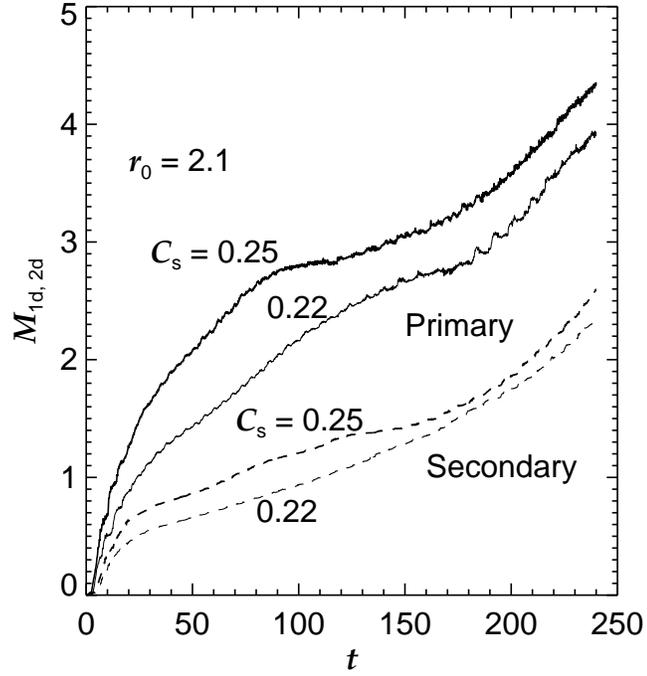}
\caption{The same as Fig.~\ref{Cs-mass} but for
comparison model 2 (thin) and 5 (thick). \label{Cs-mass2}}
\end{figure}

Figure~\ref{Cs-mass2} is the same as Figure~\ref{Cs-mass}
but for comparison between models 2 and 5 in which only
the sound speed, $ c _s $, is different.  The disk masses
are $\sim $30 \% larger in model 5.
When the gas temperature is higher, the accretion rate
is higher.  The gas temperature affects a little the 
ratio between the primary and secondary accretion rates. 

\begin{figure}
\centering
\includegraphics[width=0.6\textwidth]{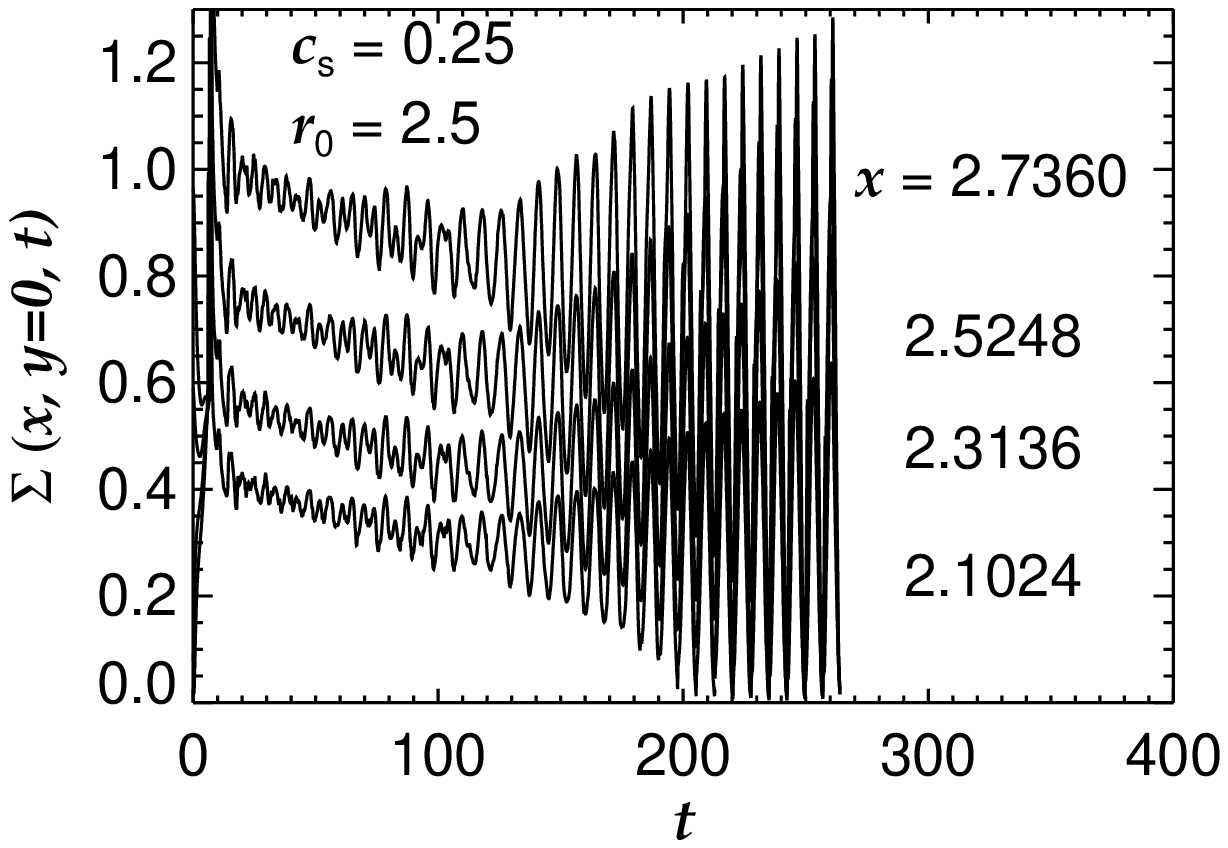}
\caption{The same as Fig.~\ref{model1-wavex} but
for model 4.\label{model3-wavex}}
\end{figure}

The vibration of the spiral shock wave is also seen
in model 3.  Figure~\ref{model3-wavex} is the same as
Figure~\ref{model1-wavex} but for model 3.  The
oscillation frequency is almost the same as that in
model 1.

\subsection{Model 6}

We have constructed model 6 to examine the effect
of the softening radius.  The initial state is
the same as that of model 3 except for  
$ \beta _1 $ and $ \beta _2 $.  The gravity is
artificially reduced in the regions of
$ |\mbox{\boldmath$r$} - \mbox{\boldmath$r$}_1 | 
< 0.14 $ and 
$ |\mbox{\boldmath$r$} - \mbox{\boldmath$r$}_2 | 
< 0.14 $ in model 6.  

\begin{figure}
\centering
\includegraphics[width=0.5\textwidth]{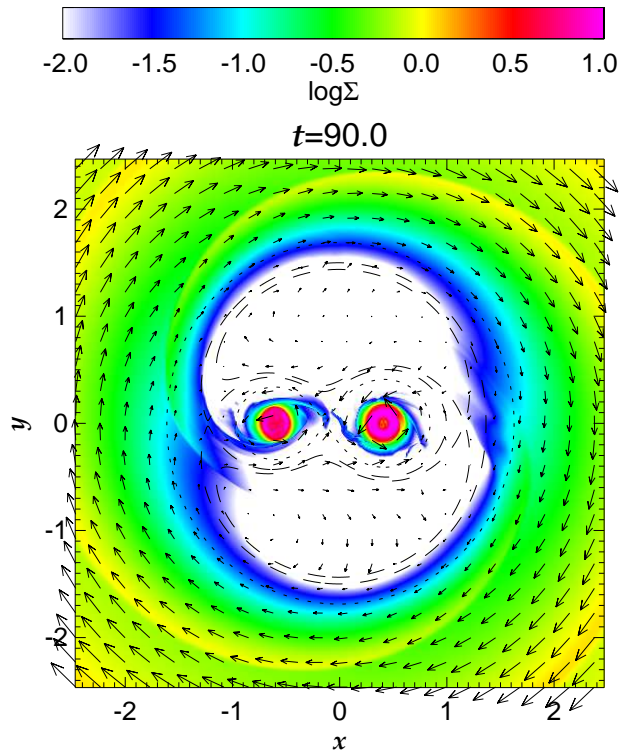}
\caption{The surface density distribution at $ t $~=~90.0
in model 6. The notation is the same as that of
Fig.~\ref{model1}. \label{model6}}
\end{figure}

Figure~\ref{model6} shows the surface density distribution
at $ t $~=~90.0 in model 6.  Also in model 6, the
the primary disk is more massive than the secondary
one.  Since the gravity is artificially reduced
in a larger area, the primary and secondary disks
are rings with larger holes in the early stages. It takes
a longer time for the gas rings to spread into 
the center. 

\begin{figure}[!hp]
\centering
\includegraphics[width=0.6\textwidth]{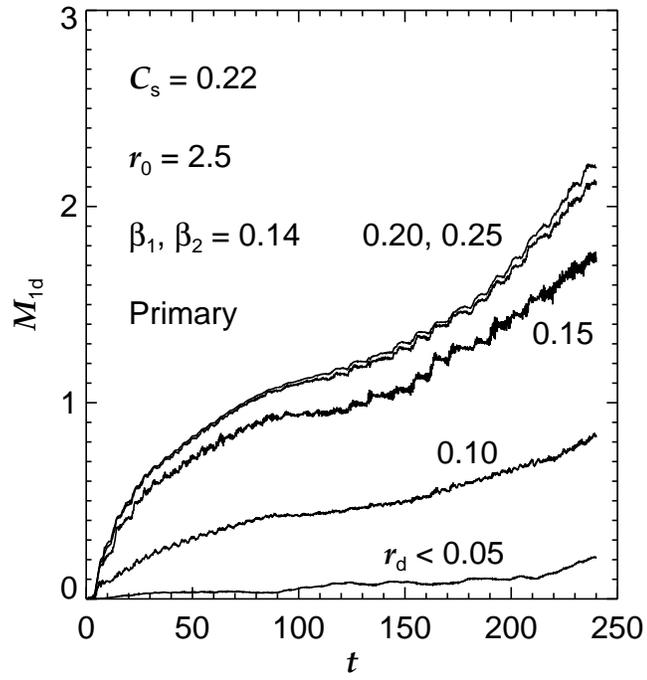}
\caption{The same ad Fig.~\ref{model1-M1} but for 
the primary in model 6. \label{model6-M1}}
\end{figure}

\begin{figure}[!hp]
\centering
\includegraphics[width=0.6\textwidth]{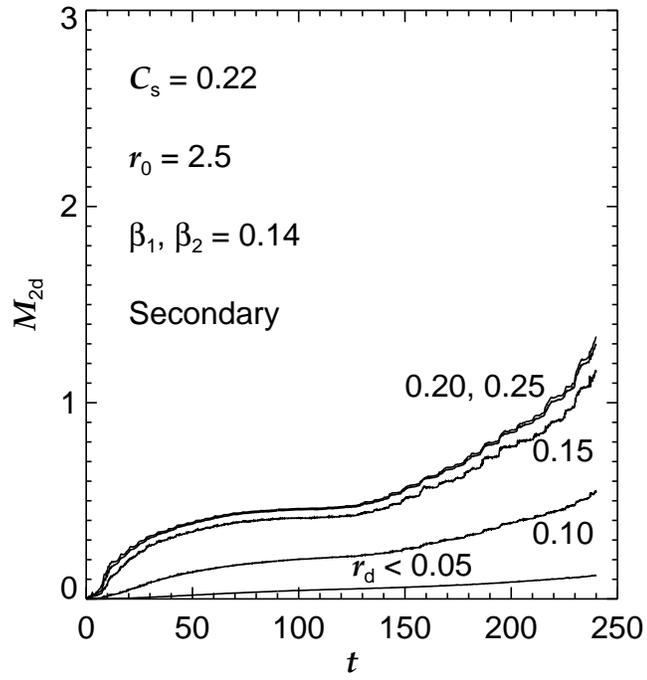}
\caption{The same ad Fig.~\ref{model1-M1} but for 
the secondary in model 6. \label{model6-M2}}
\end{figure}

Figures~\ref{model6-M1} and \ref{model6-M2} are
the same as Figures~\ref{model1-M1} and ~\ref{model1-M2} 
but for the
primary and secondary disks in model 6, respectively.
The mass of the secondary disk is smaller than
in model 3. Note, however, that the softening
radius is still as small as $ \beta _2 $~=~0.14.
The secondary disk extends outside the softening
radius and can capture gas if it is accreted.
The primary disk sheds its outer part at some
epochs (e.g. $ t \simeq 150 $ and 190) and 
the secondary receives it.
This feature is also seen in model 6.  

The softening radius affects the inner structure of
the circumstellar disks.  However, it does not
affect the gas stream from L$_2$ point directly;
the gas flows far from the region of artificial
reduction in the gravity.  We conclude that the
softening radius does not give a serious effect on the
result that the primary accretes more than the
secondary.

\subsection{Model 7}

We have constructed model 7 to examine the effect
of initial surface density distribution.  Model 7
is the same model 2 except for $ h $ which denotes
the surface density gradient at the initial disk edge.
The surface density increases gradually in model 7
since $ h~=~0.6$.  Hence the initial pressure force is
much weaker than in model 2, since $ h~=~0.144 $ in model 2.

\begin{figure}[!hp]
\centering
\includegraphics[width=0.6\textwidth]{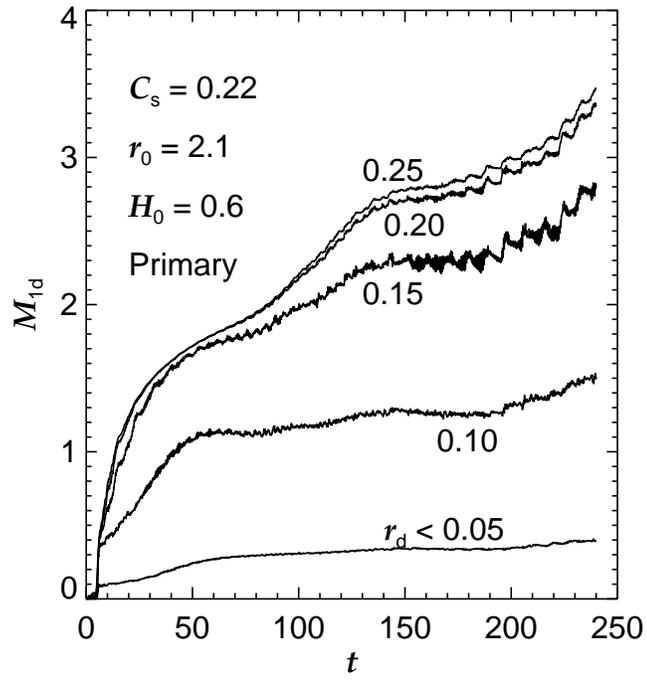}
\caption{The same ad Fig.~\ref{model2-M1} but for 
that in model 7. \label{model7-M1}}
\end{figure}

\begin{figure}[!hp]
\centering
\includegraphics[width=0.6\textwidth]{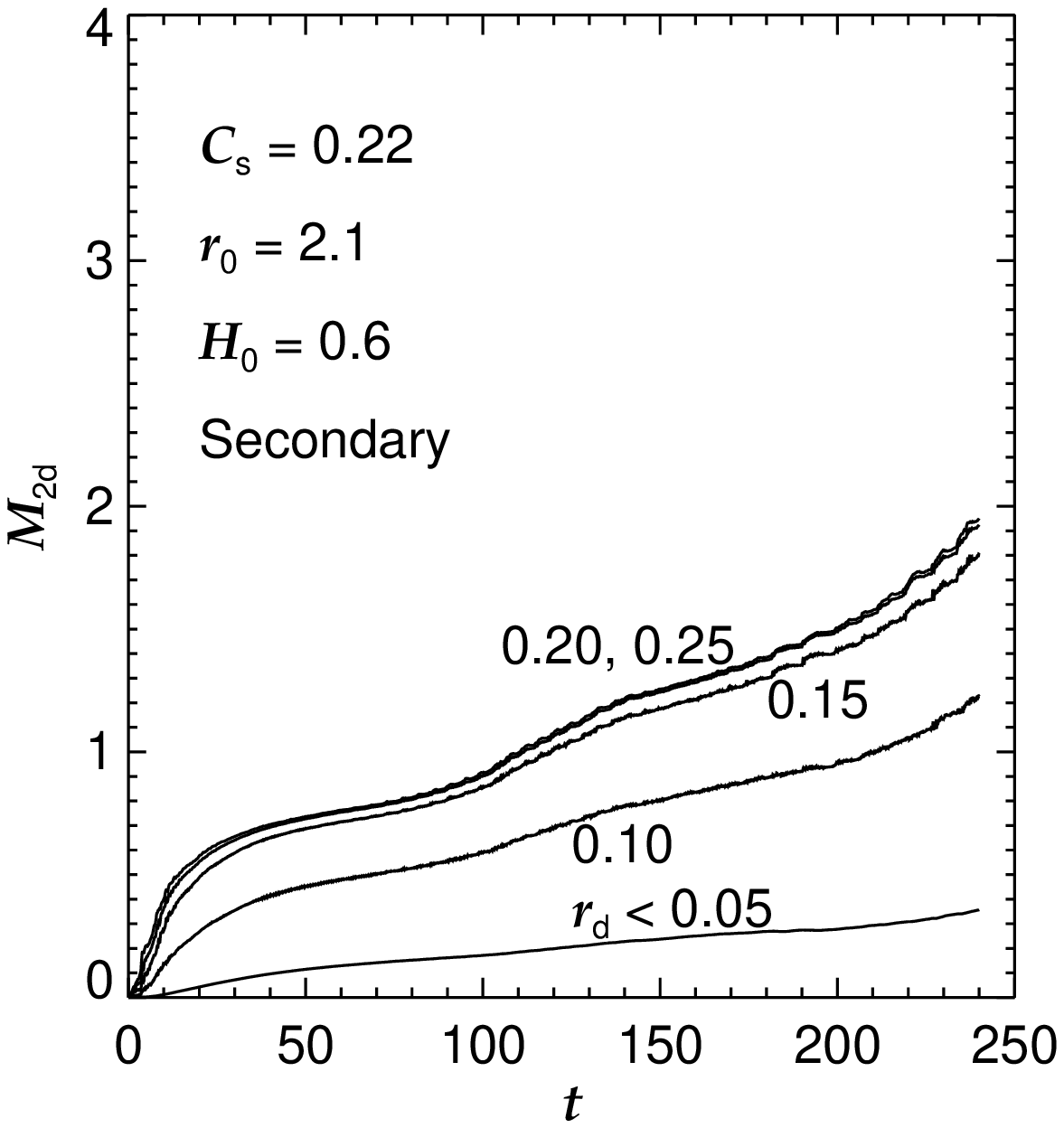}
\caption{The same ad Fig.~\ref{model2-M2} but for 
model 7. \label{model7-M2}}
\end{figure}

Figures~\ref{model7-M1} and \ref{model7-M2} show the
disk mass as a function of time in model 7 
for primary and secondy, respectively.  The initial
rise in the primary disk mass is steeper in model 7.
This is because the initial disk is extended more
in model 7.  Inner part of the circumbirnary disk is 
truncated even when the pressure gradient is weak.
The primary disk mass of model 7 is almost the same
as that of model 2.  The disk mass does not depend
seriously on $ h $.

\section{DISCUSSIONS}

In this section, we discuss the softening radius, 
the hot spot, the accretion impact on
the circumstellar disk, and the oscillation of spiral
waves excited in the circumbinary disk.

In this work the softening radii, $\beta _1 $ and $ \beta _2 $,
are three times smaller than those in our previous work 
(Paper I).  They are also much smaller than the circumstellar
disk radii. Our numerical simulations clearly resolved internal 
structures of the circumstellar disks such as spiral waves within
them.  The simulations still depend a little on the softening
radius.  The gas accretion is slow inside the softening radii
and the circumstellar disks have inner holes in the early period.
The circumstellar disks affect accretion rates through dynamical
interaction with the gas stream from L2 point.  

Further reduction in the softening radius requires 
extremely high spatial resolution and accordingly
unfeasible computation cost. See Appendix 
on the spatial resolution required.

As shown in the previous section, the primary disk 
has a hot spot during the primary accretion rate
dominates over the secondary one.  The hot spot
is due to the impact of the gas stream and the
gas looses its kinetic energy through shock
dissipation.  The kinetic energy dissipation
plays an important role for a gas element to
be accreted onto a circumstellar disk.  

\begin{figure}
\centering
\includegraphics[width=0.5\textwidth]{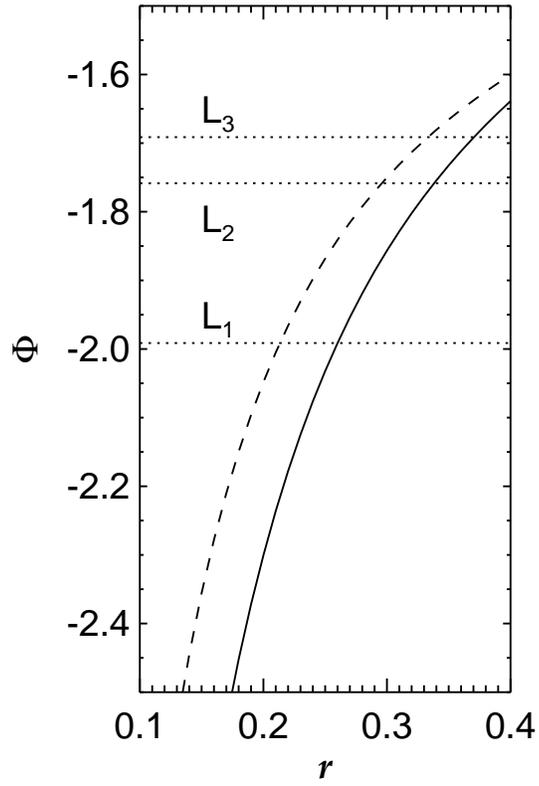}
\caption{The Jacobi integral is shown for 
circular orbits around the primary (solid) and
secondary (dashed).  The abscissa is the orbital
radius.  The value of Jacobi integral is also shwon
for particles at rest on the Lagrangian points,
L$_1$, L$_2$, and L$_3$ for reference. \label{Bernoulli}}
\end{figure}

It is well known that the Jacobi integral,
\begin{equation}
J \; = \; \frac{|\mbox{\boldmath$u$}| ^2}{2} \; + \;
\Phi \, ,
\end{equation}
is a constant of motion for a particle moving around
a circular binary.  After some algebraic manipulation,
we obtain
\begin{equation}
\frac{D J^\prime}{Dt} \; = \; - \, c _s ^2 \,
\frac{\partial \ln \Sigma}{\partial t} \, 
\label{Jacobi}
\end{equation}
for an isothermal gas flow without shock, where
\begin{equation}
J ^\prime \; = \; J \; + \; c _s ^2 \, \ln \Sigma .
\end{equation}
The difference between $ J $ and $ J ^\prime $ is
small when the gas temperature is low.  The right
hand side of Equation (\ref{Jacobi}) is also small.
Thus the Jacobi integral can be regarded as 
a constant for a Lagrangian gas element if
the gas is cold.  When a particle orbits around the
primary at the Keplerian velocity, the Jacobi integral
is evaluated to be
\begin{equation}
J _1 \; = \; - \, M _2 \, - \, \frac{M_2 ^2}{2} \,
+ \, \frac{1}{2} \left( \sqrt{\displaystyle\frac{M _1}
{|\mbox{\boldmath$r$} \, - \, \mbox{\boldmath$r$} _1|}} 
\, - \, |\mbox{\boldmath$r$} \, - \, \mbox{\boldmath$r$} _1| 
\right) ^2 \, ,
\end{equation}
where the tidal force due to the secondary is neglected 
for simplicity.  Similarly, it is evaluated to be 
\begin{equation}
J _2 \; = \; - \, M _1 \, - \, \frac{M_1 ^2}{2} \,
+ \, \frac{1}{2} \left( \sqrt{\displaystyle\frac{M _2}
{|\mbox{\boldmath$r$} \, - \, \mbox{\boldmath$r$} _2|}} 
\, - \, |\mbox{\boldmath$r$} \, - \, \mbox{\boldmath$r$} _2| 
\right) ^2 \, ,
\end{equation}
for a particle orbiting around the secondary.
The values of $ J _1 $ and $ J _2 $ are shown 
as a function of the orbital radius in
Figure~\ref{Bernoulli}.  
The dashed lines denote 
the potential energy at the Lagrangian points
L$_1$, L$_2$, and L$_3$.

The inflowing gas has a small velocity in the 
corotation frame when it passes L$_2$ point.  
Thus the Jacobi integral is nearly equal to the
potential energy at L$_2$ point.  Provided that
the gas accretes onto either the primary or the
secondary without changing the Jacobi integral.  
Then the orbital radius should be as much as 
$ | \mbox{\boldmath$r$} \, - \, 
\mbox{\boldmath$r$} _1 |$~=~0.34 or
$ | \mbox{\boldmath$r$} \, - \, 
\mbox{\boldmath$r$} _2 |$~=~0.30.
The orbital radius can be smaller only when the
Jacobi integral is reduced substantially through
strong shock dissipation.   

When the orbital radius is $ | \mbox{\boldmath$r$} \, - \, 
\mbox{\boldmath$r$} _1 |$~=~0.26, the Jacobi integral,
$ J _1 $, is equal to that at rest at L$_1$ point.
This implies that a gas element is marginally bound to
the primary when the orbital radius is 0.26. 
The gas element may flows into the secondary lobe
if perturbed a little.  Our simulations indeed show
such mass transfer from the primary disk to the secondary
one around $ t~\simeq~190 $ in model 1.  
The outer part of primary disk is shed into 
the secondary. The radius of the marginally
bound orbit is $ | \mbox{\boldmath$r$} \, - \, 
\mbox{\boldmath$r$} _2 |$~=~0.21 for the secondary
disk.  The secondary disk is confined mostly inside 
the marginal orbit in our simulations. 

The above discussion gives a guideline for choosing the
softening radius.  A gas element orbiting at the softening
radius should have an appreciably smaller Jacobi integral 
than resting at L1 point.  If the gas element has an 
eccentric orbit around a star, the pericenter should be 
located outside the softening radius.  Since the softening
radius was 20 \% of the binary separation in Paper I, 
it is close to the radius of the marginally bound orbit
for the secondary and should be set smaller.

The kinetic energy dissipation may brighten 
the hot spot at some wavelengths.  Although the 
hot spot is assumed to be isothermal in our simulations
for simplicity, it should be hotter than the rest part of
the circumstellar disk. It should be analogous to the hot spot
in the cataclysmic variables and X-ray binaries 
\citep[see, e.g.,][]{marsh01}. In these 
systems, a compact star (either of white dwarf, neutron star
and black hole) accretes gas from its companion through
L$_1$ point. The gas inflow forms a hot spot on the circumstellar
disk around the compact star.  It is often identified by
occultation.  Similarly, the hot spot may 
be observed in binaries of young stellar objects.
It may be identified by the Doppler effect if it is luminous
at some emission lines. 

It should be noted that the impact depends on the 
circumprimary disk.  When the circumprimary disk has a 
higher surface density and pressure, the inflowing gas 
stream is relatively weaker and hence produces the
impact point at an outer radius.  The evolution of 
circumstellar disks is controlled by angular momentum
transfer and hence will be affected by assumed viscosity 
and gravity softening. 

Next we discuss the oscillation of spiral waves excited
in the circumbinary disk. The oscillation is closely
related to gas accretion from the circumbinary disk
as shown in the previous section.  This is also confirmed
by Equation (\ref{Jacobi}).  If the density distribution 
is stationary in the corotation frame, the modified Jacobi 
constant, $ J ^\prime $, is constant except when it 
gets across a shock front.  The shock dissipates the kinetic
energy and decreases the Jacobi constant.  If the Jacobi
constant decreases, the orbital radius increases.
This is because  
the Jacobi constant is evaluated to be
\begin{equation}
J \; = \; - \, \frac{1}{2r} \, - \, \sqrt{r} \, ,
\label{Jacobi-b}
\end{equation} 
when the gas element rotates around the binary with the
orbital radius, $ r $, in the circumbinary disk.
Equation (\ref{Jacobi-b}) is derived under the assumption
that the orbit is almost circular and the velocity is
Keplerian.  The Jacobi constant is larger for a larger
$ r $ when $ r \, > \, 1$.  In other words, the gas element
gains angular momentum from the spiral wave corotating
with the binary, since the angular velocity of the binary
is higher than that of a gas element in the circumbinary disk.
This means that a stationary shock wave drives the circumbinary
disk outward.  The accretion is due to vibration of the shock wave.  
Some gas elements move inward by loosing the angular momentum 
and Jacobi constants through exchange with some other gas
elements, when the spiral shock wave is not stationary.

The oscillation has a typical period of $\sim 4.5$ when
measured in the corotation frame.  Accordingly the wave
has an apparent phase velocity of $\sim 0.7 $ in the
corotation frame since the azimuthal wavenumber is $ m = 2 $. 
Then the phase velocity is $ \sim 0.3 $ in the rest frame
and coincides with the angular velocity of gas orbiting
around the binary at $ r \, \simeq \, 2.2 $. This 
implies that the wave is excited by resonance at the
inner edge of the circumbinary disk. 

The amplitude of the oscillation varies on the timescale
of several tens rotation period.  The variation is coherent
in the radial direction.  When the amplitude is small, the
whole cicumbinary disk is nearly stationary. 
When the amplitude is large, both the radial and azimuthal
components of velocity oscillate at the same frequency. 
The variation may induce variability in the accretion onto
the binary.  

\acknowledgments

We thank Tomoaki Matsumoto for his useful comments and advices.
This study is financially supported in part by the 
Grant-in-Aid for Scientific Research on Priority Area (19015003) of 
The Ministry of Education, Culture, Sports, Science, and
Technology (MEXT).

\appendix

\section{REFERENCE VELOCITY METHOD}

We describe the reference velocity methods for integrating 
hydrodynamical equations.

First we divide the computation domain into four areas:
one around the primary, another around the secondary, 
another surrounding the binary, and the rest.  The
reference velocity is defined differently in each area.

The reference velocity is defined to be the Keplerian rotation around the 
center of mass,
\begin{equation}
\mbox{\boldmath$w$} \; = \; \omega (|\mbox{\boldmath$r$}|) \,
\mbox{\boldmath$e$} _z \times \mbox{\boldmath$r$} 
\end{equation}
\begin{equation}
\omega (r) =  -  1 + r ^{-3/2} ,
\end{equation}
outside the circle of $ |\mbox{\boldmath$r$} | \ge 1  $.

The reference velocity is taken to be circular rotation around
the primary,
\begin{equation}
\mbox{\boldmath$w$} \; = \; \omega (|\mbox{\boldmath$r$} \, - \,
\mbox{\boldmath$r$} _1 |) \,
\mbox{\boldmath$e$} _z \times \left(
\mbox{\boldmath$r$} \, - \, \mbox{\boldmath$r$} _1 \right) \,
\end{equation}
in the vicinity of $ |\mbox{\boldmath$r$} \, - \, 
\mbox{\boldmath$r$}_1 | \, \le \, \gamma _1 $.
The angular velocity is set to be
\begin{equation}
\omega \left(|\mbox{\boldmath$r$} \, - \, \mbox{\boldmath$r$} _1 | 
\right) \; = \; \left\{
\begin{array}{ll} 
- \, \Omega \, + \, \displaystyle \left( \frac{M_1}{\beta _1 ^3}
\right) ^{1/2} \,
\left[ \frac{7}{4} \, - \, \frac{3}{4} \, 
\left(\frac{|\mbox{\boldmath$r$} \, - \, \mbox{\boldmath$r$} _1 |}
{\beta _2}\right) ^2 \right]
& \left( |\mbox{\boldmath$r$} \, - \, \mbox{\boldmath$r$} _1 | 
\, < \, \beta _1 \right)
\\ - \, \Omega \, + \, \displaystyle \left(
\frac{M _1}{|\mbox{\boldmath$r$} \, - \, \mbox{\boldmath$r$} _1 | ^3} 
\right) ^{1/2} &
\left( \beta _1 \, \le \, 
|\mbox{\boldmath$r$} \, - \, \mbox{\boldmath$r$}_1 |
\, \le \, \alpha _1 \right)
\\ - \, \Omega \, + \, \displaystyle \left(
\frac{M _1}{\alpha _1 ^3} \right) ^{1/2} \,
\left( \frac{5}{2} \, - \, \frac{3}{2} 
\frac{|\mbox{\boldmath$r$} \, - \, \mbox{\boldmath$r$} _1 |}
{\alpha _1} \right) &
\left(\alpha _1 \, < \, |\mbox{\boldmath$r$} \, - \, \mbox{\boldmath$r$} _1 |
\, < \, \gamma _1\right)
\end{array}
\right. .
\end{equation}
The angular velocity coincides with the Keplerian velocity
around the primary in the region of
$ \beta _1 \, \le \, r _1 \, \le \, \alpha _1 $.  
It is slower than the Keplerian velocity in the region
of $ \alpha _1 \, \le \, r _1 \, \le \, \gamma _1 $.
The radius, $ \gamma _1 $, is set 
to be
\begin{equation}
\gamma _1 \; = \; \frac{2 \alpha _1}{3} \,
\left[ \frac{5}{2} \, \left(\frac{M _1}{\alpha _1 ^3} \right)
\, - \, \Omega \right] \, ,  
\end{equation}
so that the angular velocity vanishes at 
$ |\mbox{\boldmath$r$} \, - \, \mbox{\boldmath$r$}_1 | 
\, = \, \gamma _1 $.

The gravity of the primary is softened so that it 
balances with the centrifugal force due to the reference 
velocity.  Thus the gravity due to the primary is reduced
to 
\begin{equation}
\mbox{\boldmath$g$} \left( 
|\mbox{\boldmath$r$} \, - \, \mbox{\boldmath$r$}_1 | \right)
\; = \; - \, \frac{M _1}{\beta _1 {}^3} \, 
\left(\mbox{\boldmath$r$} \, - \, \mbox{\boldmath$r$}_1 \right) \,
\left[ \frac{7}{4} \, - \, 
\frac{|\mbox{\boldmath$r$} \, - \, \mbox{\boldmath$r$}_1 |}
{\beta _2 {}^2} \right] \, .
\end{equation}

Similarly the reference velocity is taken to be circular 
rotation around the secondary in the vicinity of the
secondary.  It is expressed as
\begin{equation}
\mbox{\boldmath$w$} \; = \; \omega (|\mbox{\boldmath$r$} \, - \,
\mbox{\boldmath$r$} _2 |) \,
\mbox{\boldmath$e$} _z \times \left(
\mbox{\boldmath$r$} \, - \, \mbox{\boldmath$r$} _2 \right) 
\end{equation}
\begin{equation}
\omega (|\mbox{\boldmath$r$} \, - \, \mbox{\boldmath$r$} _2|) 
\; = \; \left\{
\begin{array}{ll} 
- \, \Omega \, + \, \displaystyle \left( \frac{M_2}{\beta _2 ^3}
\right) ^{1/2} \,
\left[ \frac{7}{4} \, - \, \frac{3}{4} \, 
\left(\frac{|\mbox{\boldmath$r$} \, - \, \mbox{\boldmath$r$} _2|}
{\beta _2}\right) ^2 \right]
& (|\mbox{\boldmath$r$} \, - \, \mbox{\boldmath$r$} _2| \, < \, \beta _2)
\\ - \, \Omega \, + \, \displaystyle \left(
\frac{M _2}{|\mbox{\boldmath$r$} \, - \, \mbox{\boldmath$r$} _2| ^3} 
\right) ^{1/2} &
(\beta _2 \, \le \, |\mbox{\boldmath$r$} \, - \, \mbox{\boldmath$r$} _2|
\, \le \, \alpha _2)
\\ - \, \Omega \, + \, \displaystyle \left(
\frac{M _2}{\alpha _2 ^3} \right) ^{1/2} \,
\left( \frac{5}{2} \, - \, \frac{3}{2} 
\frac{|\mbox{\boldmath$r$} \, - \, \mbox{\boldmath$r$} _2|}
{\alpha _2} \right) &
(\alpha _2 \, < \, |\mbox{\boldmath$r$} \, - \, \mbox{\boldmath$r$} _2|
 \, < \, \gamma _2)
\end{array}
\right. \,
\end{equation}
where
\begin{equation}
\gamma _2 \; = \; \frac{2 \alpha _2}{3} \,
\left[ \frac{5}{2} \, \left(\frac{M _2}{\alpha _2 ^3} \right)
\, - \, \Omega \right] \, ,  
\end{equation}
The values of $ \alpha _i $, $ \beta _i $, and $ \gamma _i $
are tabulated in Table~\ref{beta}. Note 
$ \gamma _1 \, + \, \gamma _2 \, < \, 1 $ and accordingly 
that the region of $ |\mbox{\boldmath$r$} \, - \, 
\mbox{\boldmath$r$}_1 | \, < \, \gamma _1 $ does not
overlap with that of $ |\mbox{\boldmath$r$} \, - \, 
\mbox{\boldmath$r$}_2 | \, < \, \gamma _2 $.

The reference velocity vanishes in the rest area, i.e.,
inside the circumbinary disk but outside the primary
and secondary disks. Remember that the reference velocity
is continuous even on the boundary between the different
areas.  Thus the derivative of the reference velocity
is regular and continuous except discontinuity on the 
boundary between the areas.  

The discontinuity of the reference velocity derivative 
has no serious 
effect on the evaluation of the source term.  
The reference velocity is analogous to the vector
potential in the electromagnetism.  Thus its derivative
is analogous to the magnetic field.  The magnetic field
can be discontinuous so the derivative of the the
reference velocity can be.  Remember that the reference
velocity is divergence free, $ \mbox{\boldmath$\nabla$}\cdot
\mbox{\boldmath$w$}\, = \, 0$.  This is analogous to the
Coulomb gauge in the electromagnetism.

The numerical flux is evaluated at the cell surface as
in the ordinary finite volume method.  The reference velocity
is evaluated analytically at the cell surface while the 
other variables are evaluated from the cell center values
by using interpolation with the minmod limiter.
Then the reference velocity is numerical viscosity free 
and the numerical solution has a larger Reynolds number.

Introduction of the reference velocity improves
evaluation of the centrifugal force greatly.  
The centrifugal force, $ \mbox{\boldmath$v\cdot\nabla v$}$,
contains a spatial derivative of the velocity.
The spatial derivative is evaluated from the spatial
difference in an ordinary finite difference method and
truncation error is inevitable in its numerical 
evaluation.  The truncation error should  be small
so that minor forces such as pressure force are
appreciated properly.  Note that the numerical derivative
is only the first order accurate near shock 
fronts in a TVD scheme \citep[see, e.g.,]{toro09}.  
When and only when the spatial 
resolution, $\Delta x $, is much smaller than
the the scaleheight,
\begin{equation}
\Delta x \; \ll \; \frac{c _s ^2 \, r^2}{GM} \,
\end{equation}
the truncation error is ensured to be much smaller than the
pressure force.  This condition is equivalent to 
\begin{equation}
\Delta x \; \ll \; \left(\frac{c _s}{v _K} \right) ^2 \, r \; ,
\label{g-condition}
\end{equation}
where $ v _K $ denotes the Keplerian velocity.

The reference velocity relaxes the above mentioned 
requirement on the truncation error.  Alternatively,
the resolution should be so small that 
the difference in reference velocity is smaller than 
the sound speed between any adjacent cells.  Otherwise
a gas element would be accelerated (or decelerated) 
too much by advection from a cell to another.  
This requirement is less tight, since it can be
expressed as
\begin{equation}
\Delta x \; \ll \; \left( \frac{c _s}{v _K}\right) \, r \; .
\end{equation}
Equation (\ref{g-condition}) is much more tight than this,
when the Mach number ($ v _K/c _s$) is large. 
The Mach number is as high as 
\begin{equation}
\frac{v _K}{c _s} \; = \; 13.3
\left( \frac{M _1}{0.596} \right) ^{1/2} \,
\left( \frac{|\mbox{\boldmath$r$} \, - \,
\mbox{\boldmath$r$} _1 |}{0.07} \right) ^{-1/2} \,
\left( \frac{c _s}{0.22} \right) ^{-1} \, 
\end{equation}
for the Keplerian rotation around the primary.
Thus the reference velocity.  Further reduction in 
the softening radius requires extremely high spatial
resolution even when the reference velocity is
introduced.  

Our numerical scheme is similar to FARGO 
(Fast Advection in Rotating Gaseous Objects) developed by 
\citet{masset00} for solving hydrodynamics a gaseous 
disk rotating around a star on fixed polar grids. 
FARGO decomposes the rotation velocity into two components,
the azimuthally averaged rotation and deviation from it. The 
advection due to the former is treated separately in FARGO.
The separation relaxes the ordinary Courant condition and speeds up 
the computation.  At the same time, numerical diffusivity is much
smaller in FARGO thanks to the two step advection.  The centrifugal
force is dominated by that due to the averaged rotation, $ J ^2/r ^3 $, 
where $ J $ denotes the specific angular momentum.  Thus 
the truncation error is reduced also in FARGO.

The reference velocity is, however, different from FARGO.
Application of FARGO is limited to the case that the grid
is set almost parallel to the gas flow.¡¡The reference 
velocity method can be applied to any numerical grids.
FARGO can take a long time step while the reference velocity
method can not.

Our numerical scheme is also similar to that of \citet{leveque98}.
He proposed a scheme to take account of gradient of a physical 
variable within a numerical cell for elliminating source terms
apparently.  In case of hydrodynamics, the gravity can be taken into
account if the pressure has different values at the
cell surfaces opposing each other. Our reference velocity
has different values at the cell surfaces to incorporate
the gravity.  His scheme is effective for a flow in 
a quasi-static balance while ours is effective for a flow
supported mainly by the centrifugal force.

%-------------------------

\clearpage

\clearpage

\clearpage

\begin{table}
\begin{center}
\begin{tabular}{lcccc}
\tableline
No. & $ c _s $ & $ r _0 $ & $ H $ & 
$ \beta _1~(=~\beta _2)$ \\
\tableline
1 & 0.22 & 2.5 & 0.144 & 0.07 \\
2 & 0.22 & 2.1 & 0.144 & 0.07 \\
3 & 0.25 & 2.5 & 0.144 & 0.07 \\
4 & 0.25 & 2.3 & 0.144 & 0.07 \\ 
5 & 0.25 & 2.1 & 0.144 & 0.07 \\
6 & 0.22 & 2.5 & 0.144 & 0.14 \\ 
7 & 0.22 & 2.1 & 0.600 & 0.07 \\
\tableline
\end{tabular}
\end{center}
\caption{Summary of models.}
\label{model}
\end{table}

\begin{table}
\begin{center}
\begin{tabular}{lrr}
\tableline
$ i $ & 1 & 2 \\
\tableline
$ M _i $ & 0.5957 &  0.4043\\
$ \alpha _i $ & 0.4000 & 0.2700 \\
$ \gamma _i $ & 0.5793 & 0.41028 \\
\tableline
\end{tabular}
\end{center}
\caption{Model parameters for the reference velocity.\label{beta}}
\end{table}

\end{document}